\documentstyle[amsfonts,eqsecnum,preprint,aps]{revtex}

\begin{document}
\draft
\title{Collective and relative variables for a classical Klein-Gordon field\thanks{%
Talk given by M. Materassi at the XIII Jornada Cientifica Estudiantil, April 
$24\div 26$, 1997, Universitad de la Habana, Cuba.}}
\author{G.Longhi\thanks{%
e-mail: longhi@fi.infn.it} and M.Materassi\thanks{%
e-mail: materassi@pg.infn.it}}
\address{Department of Physics, University of Firenze}
\date{\today }
\maketitle

\begin{abstract}
In this paper a set of canonical collective variables is defined for a
classical Klein-Gordon field and the problem of the definition of a set of
canonical relative variables is discussed.

This last point is approached by means of a harmonic analysis in momentum
space. This analysis shows that the relative variables can be defined if
certain conditions are fulfilled by the field configurations.

These conditions are expressed by the vanishing of a set of conserved
quantities, referred to as supertranslations since as canonical observables
they generate a set of canonical transformations whose algebra is the same
as that which arises in the study of the asymptotic behaviour of the metric
of an isolated system in General Relativity \cite{bms-matelonghi}.
\end{abstract}


\newpage

\newpage

\preprint{Th-May,1997}

\bigskip

\narrowtext

\section{Introduction.}

In the search for a theory of $N$ relativistic interacting particles a
possible way is the use of constrained dynamics \cite{LEV}. This is a
phase-space approach based on a set of mass constraints, one for each
particle, in involution among themselves, or first class following Dirac 
\cite{DIR}, where the masses are potential masses depending on the relative
coordinates of the particles as well on their momenta.

Since each constraint can be interpreted as a Hamilton-Jacobi equation \cite
{DLGP} and their first class character as the corresponding integrability
condition, one describes the dynamics in terms of $N$ independent time
parameters, with no necessity of introducing gauge fixing conditions.

This approach has many conceptual advantages but, from the point of view of
model building, apart from the case of $N=2$ \cite{TWO}, shows great
difficulties if we try to fulfill the cluster decomposition (also called
separability) \cite{SEP}.

So it is desirable to have a one-time formulation, preserving the
relativistic invariance in a controlled way, where the cluster decomposition
property could be more easily satisfied.

A way to do this was suggested many years ago by Dirac \cite{DIR}. He
suggested to introduce a foliation of Minkowski spacetime by a family of
space-like surfaces; the parameter of such a family can be regarded as a
unique time parameter, and can be used for a one-time formulation of the
dynamics of an $N$-particle system. The various quantities which
characterize each space-like surface of the family become dynamical
variables, which must be added to the phase-space of the system.

A great simplification can be obtained with suitable gauge fixings which
restrict the surfaces to be space-like planes: in the case of an isolated
system there is the possibility of using the conservation of the total
momentum of the system to define, in an intrinsic way, a family of
space-like planes orthogonal to the total momentum $P_\mu $. It amounts in
using as time the one that would be measured in the rest frame of the total
momentum.

This approach has been developed by Lusanna \cite{LUSALBA}: following this
line it becomes of interest to consider fields besides particles to take
account of their interactions, for instance electromagnetic interactions.

In order to completely work out a correct canonical framework of the
isolated system of matter and fields (treated as in \cite{LUSALBA}), it
becomes necessary to define ''center of mass'' and relative variables for
the fields too: by using this canonical collective-relative variables the
complete Dirac-Bergmann \cite{Dirberg} reduction of the whole gauge system
is possible.

It is well known that the definition of a center of mass in special
relativity presents various difficulties \cite{CMRS}. It is not possible to
have a sound definition of a center of mass coordinate adapted to the first
class constraints of the model under study, which is at the same time
canonically conjugated to the total momentum and covariant. Nevertheless,
for the purpose discussed here, it can be sufficient to define ''some''
canonical variable $X^\mu $ conjugated to the momentum, renouncing for the
moment to an explicit physical meaning analogous to the Newtonian one. We
will reserve the name {\it collective coordinates} for this canonical
coordinate and for the total momentum; the other independent canonical
coordinates will be referred to as {\it relative}.

Moreover, there will be the problem of defining such variables for an
infinite-dimensional system like a field. We propose in this paper an
approach to this problem and show, in the simplest case of a real scalar
field, that it is possible to define a canonical set of collective and
relative variables with some restrictions that appear to be as integrability
conditions: our recipe for the definition of collective variables seems to
work only within a particular class of field configurations; there will be
the necessity to understand if these restrictions are of a general nature
(and are eventually met whenever one tries to define collective and relative
variables for a relativistic field), or only affect our particular way to
work out the variable $X^\mu $.

Anyway, when collective and relative variables are singled out, the
symplectic Lorentz algebra splits into the direct sum of two copies of $%
{\frak so}\left( 1,3\right) $, the first involving only relative variables,
the second depending only on $X$ and $P$; this gives rise to a decomposition
of the Poincar\'e algebra in a direct sum of a collective inhomogeneous
Lorentz algebra, and a homogeneous relative one. In this way we get a
definition of the total conserved intrinsic angular momentum of the system.

The path we have chosen for the definition of this canonical basis was the
use of a canonical modulus-phase set in momentum space. This is not of
course the only way and probably not the most convenient, indeed the
relation between the new variables and the old ones is very involved.
Nevertheless, as far as the authors know, no attempt of this kind is present
in the literature.

In Section II we discuss a possible choice of the collective canonical
variables. In order to define the variable $X^\mu $ conjugated to the total
momentum $P^\mu $, we introduce a partially arbitrary function $F$, which is
a function of $\left( P\cdot k\right) $ and $P^2$, where $k$ is the momentum
space variable of the field. This is quite analogous to the classical
particle case, where the center of mass position is defined in terms of a
set of weights, which are arbitrary apart from a normalization condition.
The transformation properties of $X^\mu $ are discussed.

In Section III we give a discussion of the relative variables, which are
required to be scalar fields. Their complete definition requires the full
analysis of the Laplace-Beltrami differential operator defined on the
mass-hyperboloid in momentum space, which has been exhaustively studied in 
\cite{bms-matelonghi}.

The Poisson algebra of the new variables with the infinitesimal generators
of Lorentz transformations is studied, and it is shown the splitting of the
Poincar\'e algebra in collective and relative parts. This decomposition was
expected in view of the analogy with the classical particle case.

Section IV is finally devoted to the discussion of the consistency
conditions of the method, that single out the field configurations for which
this all seems to work: they form a very well defined family of
configurations selected for purely mathematical reasons at this level;
anyway we think their physical content could be deeper, and it would deserve
further study.

These consistency conditions essentially involve those quantities $%
P_{l,m}\left[ \Phi ,\Pi \right] $ referred to as supertranslations (see \cite
{bms-matelonghi} and references therein), that appeared for the first time
in a general relativistic context.

The definitions and notations of the Klein-Gordon field are given in
Appendix A. In Appendix B some useful Poisson brackets are worked out, while
in Appendix C the Laplace-Beltrami operator needed by the whole study is
discussed.

In Appendix D we give some details of the calculation of the Green function
of the Laplace-Beltrami operator, which is necessary for the canonical
transformation to the new variables and for its inverse. Finally, in
Appendix E we construct the Poisson algebra of the Poincar\'e group in terms
of the new relative set.

\section{The collective variables.}

\subsection{Definition of the collective variables.}

In this Section we are interested in the search of a variable conjugated to
the total 4-momentum of the field (see Appendix A),

\begin{equation}
P^\mu =\int \tilde {dk}k^\mu \overline{a}(k)a(k).  \label{1.1}
\end{equation}
(we will use the symbol $f\left( k\right) $ for simplicity, instead of $f(%
\vec k)$).

For a field $\Phi (x)$ solution of the Klein-Gordon equation $P^\mu $ is a
timelike four-momentum \cite{JOS}. This $P^\mu $ is finite if 
\begin{equation}
\begin{array}{c}
\begin{array}{cccc}
\left| a\left( k\right) \right| \simeq \left| \vec k\right| ^{-\frac 32%
-\sigma }, & \sigma >0, & {\rm for} & \left| \vec k\right| >>m,
\end{array}
\\ 
\begin{array}{cccc}
\left| a\left( k\right) \right| \simeq \left| \vec k\right| ^{-\frac 32%
+\epsilon }, & \epsilon >0, & {\rm for} & \left| \vec k\right| <<m,
\end{array}
\end{array}
\label{1.1.bis}
\end{equation}
and this condition allows the other Poincar\'e generators (as given in (\ref
{A.18}) and (\ref{A.19})) to be finite too.

A possible approach to the problem of finding some $X$ conjugated to $P$ is
to look for a field $\phi (k)$ conjugated to the square modulus of $a(k)$

\begin{equation}
I\left( k\right) =\overline{a}\left( k\right) a\left( k\right) ,  \label{1.2}
\end{equation}

\noindent that is

\begin{equation}
\{I\left( k\right) ,\ \phi \left( k^{\prime }\right) \}=\Omega \left(
k\right) \delta ^3(\vec k-\vec k^{\prime }).  \label{1.3}
\end{equation}

The general solution for $\phi $ is the phase

\begin{equation}
\phi (k)={\frac 1{2i}}\ln \frac{a(k)}{\overline{a}(k)},  \label{1.4}
\end{equation}

\noindent (where the principal value of the logarithm is meant) plus an
arbitrary function of $I(k)$. The inverse is

\begin{equation}
a(k)=\sqrt{I(k)}\exp {i}\phi (k),  \label{1.5}
\end{equation}

\noindent and the Poisson bracket $\{I\left( k\right) ,\ \phi \left(
k^{\prime }\right) \}$ is exactly what we needed.

Relation (\ref{1.3}) has been criticized by Dubin et al. \cite{DUB}. These
authors observed that, if we interpret the square modulus $I(k)$ and the
phase $\phi (k)$ as distributions in $S^{\prime }({\bf R}^3)$ (the space of
tempered distributions on ${\bf R}^3$), due to a discontinuity in any
possible definition of the phase, one necessarily gets a non canonical term
added to the right hand side of Eq.(\ref{1.3}). This noncanonical term is a
distribution having the form of Dirac comb with support on the values of $k$
which give $\phi (k)$ equal to some integer multiple of $2\pi $, depending
on the definition of the original function $\phi $.

We will adopt the point of view that the phase is a function. In this case
only the classical derivatives of the phase will appear in the Poisson
bracket, and no anomalous term.

Let us now define a collective coordinate, conjugated to the total momentum $%
P^\mu $.

\noindent It is well known that no center of mass coordinate in the usual
sense exists in relativistic theories \cite{CMRS}; in any case we may define
a coordinate conjugated to $P^\mu $. Indeed we may choose

\begin{equation}
X^{\mu} = \int\tilde{dk}\phi(k){\frac{\partial}{{\partial P_{\mu}}}} F(P, k),
\label{1.7}
\end{equation}

\noindent where $F(P,k)$ is a scalar function of $P$ and $\vec k$, i.e. a
function of $P^2$ and $\left( P\cdot k\right) $, where $k^0=\omega (k)=\sqrt{%
\left| \vec k\right| ^2+m^2}$.

The function $F(P,k)$ is the analogous of the weights required in order to
define the center of mass position in particle physics. They undergo a
normalization condition to render the center of mass coordinate conjugate to
the total momentum.

The conditions we put on the function $F$ are:

a)-normalization:

\begin{equation}
\int\tilde{dk} k^{\mu} F(P, k) = P^{\mu},  \label{1.8}
\end{equation}

\noindent which can be satisfied by any scalar function, since the integral
on the left hand side is, by covariance, equal to $P^\mu $ times a function
of $P^2$. Dividing by this function we may identically satisfy the
normalization (\ref{1.8}).

b)-Reality:

\begin{equation}
\overline{F(P, k)} = F(P, k).  \label{1.9}
\end{equation}

c)-Integrability and differentiability. The existence of the integral in Eq.(%
\ref{1.7}) depends on the behaviour of the phase $\phi (k)$, which will be
discussed in the sequel. We will assume for $F$ the necessary behaviour for $%
\left| \vec k\right| \rightarrow \infty $ and for $\left| \vec k\right|
\rightarrow 0$.

A simple choice of $F$ could be $F\propto e^{-\left( P\cdot k\right) }$, to
be properly normalized. Since both $P^\mu $ and $k^\mu $ are time-like and
future directed, this choice provides a smoothing factor at infinity in
momentum space.

Using the Eq.(\ref{1.8}) we find

\begin{equation}
\{P^{\mu},\ X^{\nu}\} = \int\tilde{dk} k^{\mu}{\frac{\partial}{\partial
P^{\nu}}} F((P\cdot k),P^2) = \eta^{\mu\nu}.  \label{1.10}
\end{equation}

Since

\begin{equation}
\{\phi(k),\ P^{\mu}\} = - k^{\mu},  \label{1.11}
\end{equation}

\noindent we have also

\begin{equation}
\{ X^{\mu},\ X^{\nu}\} = 0.  \label{1.12}
\end{equation}

Indeed

\[
\begin{array}{l}
\{X^\mu ,\ X^\nu \}= \\ 
=%
\displaystyle \int 
\tilde {dk}k_\rho 
{\displaystyle {\partial \over \partial P_\mu }}
F((P\cdot k),P^2)%
\displaystyle \int 
\tilde {dk^{\prime }}\phi (k^{\prime })%
{\displaystyle {\partial ^2 \over \partial P_\rho \partial P_\nu }}
F((P\cdot k^{\prime }),P^2)-(\mu \leftrightarrow \nu )= \\ 
=\eta _\rho ^\mu 
\displaystyle \int 
\tilde {dk}\phi (k)%
{\displaystyle {\partial ^2 \over \partial P_\rho \partial P_\nu }}
F((P\cdot k^{\prime }),P^2)-(\mu \leftrightarrow \nu )=0.
\end{array}
\]
Let us observe that a redefinition of the phase $\phi (k)$ by

\begin{equation}
\begin{array}{cc}
\phi (k)\rightarrow \phi (k)+2\pi N(k), & N(k)\in {\bf Z},
\end{array}
\label{1.13}
\end{equation}

\noindent produces a canonical transformation on the variables $P^\mu $ and $%
X^\mu $, under which $P^\mu $ is unchanged and $X^\mu $ is transformed in

\begin{equation}
X^{\prime\mu} = X^{\mu} + {\frac{\partial}{\partial P_{\mu}}} {\cal G}(P^2),
\label{1.14}
\end{equation}

\noindent where

\begin{equation}
{\cal G}(P^2) = 2\pi\int\tilde{dk} N(k) F((P\cdot k),P^2).  \label{1.15}
\end{equation}

\subsection{Transformation properties of $P^\mu $ and $X^\mu $.}

The field $\Phi (x)$ is, by hypothesis, a scalar field, that is, if $%
U(\Lambda ,a)$ represents the Poincar\'e transformation $(\Lambda ,a)$ on
the field $\Phi $ we have

\begin{equation}
(U(\Lambda, a)\Phi)(x) = \Phi(\Lambda^{-1}(x - a)).  \label{1.16}
\end{equation}

This yields the following transformation on the modulus and phase

\begin{equation}
(U(\Lambda ,a)I)(k)=I(\Lambda ^{-1}k),\quad (U(\Lambda ,a)\phi )(k)=\phi
(\Lambda ^{-1}k)+(k\cdot a).  \label{1.17}
\end{equation}
Denoting with a prime the corresponding transformed quantities, we may
easily verify that under the transformation (\ref{1.16}) we have

\begin{equation}
P^{\prime \mu }=\Lambda ^\mu \,_\nu P^\nu ,\quad X^{\prime \mu }=\Lambda
^\mu \,_\nu X^\nu +a^\mu .  \label{1.18}
\end{equation}

Our $X$ transforms exactly as the spacetime coordinate of an event in $M_4$.

In terms of modulus and phase, the generators of the Poincar\'e algebra are

\begin{equation}
P^{\mu} = \int\tilde{dk} k^{\mu} I(k),  \label{1.19}
\end{equation}

\begin{equation}
M_{ij}=\int \tilde {dk}I(k)\left( k^i\frac \partial {\partial k^j}-k^j\frac 
\partial {\partial k^i}\right) \phi (k),  \label{1.20}
\end{equation}

\begin{equation}
\begin{array}{cc}
M_{0j}=x_0P_j+M_{0j}^{\prime }, & M_{0j}^{\prime }=-%
\displaystyle \int 
\tilde {dk}I(k)\omega (k){%
{\displaystyle {\partial \over \partial k^j}}
}\phi (k).
\end{array}
\label{1.21}
\end{equation}

For the existence of these generators, and in order to be able to perform an
integration by parts in the expression of $M_{\mu \nu }$, we have to require
the following asymptotic behaviour of the fields $I(k)$ and $\phi (k)$, as $|%
\vec k|\rightarrow \infty $ and as $|\vec k|\rightarrow 0$

\begin{equation}
\begin{array}{cccc}
|I(k)|\simeq |\vec k|^{-3-\sigma }, & |\phi (k)|\simeq |\vec k| & {\rm \ as\ 
} & |\vec k|\rightarrow \infty ,
\end{array}
\label{1.22}
\end{equation}
\begin{equation}
\begin{array}{ccccc}
|I(k)|\simeq |\vec k|^{-3+\epsilon }, & |\phi (k)|\simeq |\vec k|^\tau , & 
\tau >-\epsilon & {\rm \ as\ } & |\vec k|\rightarrow 0,
\end{array}
\label{1.23.bis}
\end{equation}
for any $\sigma ,\epsilon >0$, which are coherent with (\ref{1.1.bis}), so
that the behaviour of $I(k)$ and $\phi (k)$ (\ref{1.22}) and (\ref{1.23.bis}%
) is consistent with the existence of the generators of the Poincar\'e
generators (\ref{1.19}), (\ref{1.20}) and (\ref{1.21}). The asymptotic
behaviour of $\phi (k)$ is determined by the transformation rule Eq.(\ref
{1.17}), while around $\left| \vec k\right| =0$ the behaviour (\ref{1.23.bis}%
) is worked out in order to keep $M_{ij}$ and $M_{0j}$ finite.

Under the action of the Poincar\'e generators the modulus and phase
transform as

\begin{equation}
\{P^{\mu},\ I(k)\} = 0,  \label{1.24}
\end{equation}

\begin{equation}
\{P^{\mu},\ \phi(k)\} = k^{\mu},  \label{1.25}
\end{equation}

\begin{equation}
\left\{ M^{ij},I(k)\right\} =\left( k^i\frac \partial {\partial k^j}-k^j%
\frac \partial {\partial k^i}\right) I(k),  \label{1.26}
\end{equation}

\begin{equation}
\left\{ M^{ij},\phi (k)\right\} =\left( k^i\frac \partial {\partial k^j}-k^j%
\frac \partial {\partial k^i}\right) \phi (k),  \label{1.27}
\end{equation}

\begin{equation}
\left\{ M^{0j},I(k)\right\} =\omega (k)\frac \partial {\partial k^j}I(k),
\label{1.28}
\end{equation}

\begin{equation}
\left\{ M^{0j},\phi (k)\right\} =\omega (k)\frac \partial {\partial k^j}\phi
(k),  \label{1.29}
\end{equation}

\noindent which can be collected in the form

\begin{equation}
\left\{ M_{\mu \nu },I(k)\right\} =D_{\mu \nu }I(k),  \label{1.30}
\end{equation}

\begin{equation}
\left\{ M_{\mu \nu },\phi (k)\right\} =D_{\mu \nu }\phi (k),  \label{1.311}
\end{equation}

\noindent where

\begin{equation}
D^{\mu \nu }=\left( \eta ^{h\mu }k^\nu -\eta ^{h\nu }k^\mu \right) \frac 
\partial {\partial k^h},  \label{1.32}
\end{equation}

\noindent and $k_0=\omega (k)$.

The differential operator $D_{\mu \nu }$ satisfies the following ${\frak so}%
\left( 1,3\right) $ algebra

\begin{equation}
\left[ D_{\mu \nu },\ D_{\rho \lambda }\right] =\eta _{\mu \rho }D_{\nu
\lambda }+\eta _{\nu \lambda }D_{\mu \rho }-\eta _{\mu \lambda }D_{\nu \rho
}-\eta _{\nu \rho }D_{\mu \lambda }.  \label{1.33}
\end{equation}

In order to find the transformation properties of the collective variables $%
P^{\mu}$ and $X^{\mu}$ under the action of the Poincar\'e's generators, we
must before develop two relations involving the function $F((P\cdot k), P^2)$%
.

Let us call $F_{/1}$ and $F_{/2}$ the derivatives of the function $F$ with
respect to the first and second argument respectively. From the form of $F$
we get

\begin{equation}
\frac \partial {\partial k^i}F=\left( P_i-\frac{k_iP_0}{\omega (k)}\right)
F_{/1},  \label{1.34}
\end{equation}

\begin{equation}
\frac \partial {\partial P^i}F=k_iF_{/1}+2P_iF_{/2},  \label{1.35}
\end{equation}

\begin{equation}
\frac \partial {\partial P^0}F=\omega (k)F_{/1}+2P^0F_{/2}.  \label{1.36}
\end{equation}

From these relations it's easy to get

\begin{equation}
\left( k_j\frac \partial {\partial k^i}-k_i\frac \partial {\partial k^j}%
\right) F=\left( P_i\frac \partial {\partial P^j}-P_j\frac \partial {%
\partial P^i}\right) F,  \label{1.37}
\end{equation}

\noindent and

\begin{equation}
\omega (k)\frac \partial {\partial k^j}F=\left( P_j\frac \partial {\partial
P^0}-P^0\frac \partial {\partial P^j}\right) F.  \label{1.38}
\end{equation}

With the use of these equations it can be shown (see Appendix B) that

\begin{equation}
\{M_{\mu \nu }^{\prime },\ X_\rho \}=-\left( \eta _{\mu \rho }X_\nu -\eta
_{\nu \rho }X_\mu \right) .  \label{1.39}
\end{equation}
In conclusion we have shown how it is possible to define a set of collective
variables. In particular the variable $X^\mu $ transforms as a four-vector,
as stated in equation (\ref{1.18}).

\section{The relative variables.}

In this Section we will give a discussion of the problem of defining a set
of relative variables. In order to have a guiding line for such search we
will take a suggestion from the example of the relativistic string model,
which was discussed from the point of view of a canonical formulation in 
\cite{CLL}. We shall need the mathematical framework that has been developed
in \cite{bms-matelonghi}; we will see that a set of canonical relative
variables can be defined at the expense of certain restrictions on the field
configuration.

\subsection{Subsidiary variables.}

We first introduce a set of subsidiary variables $\hat I(k)$ and ${\hat \phi 
}(k)$ such that

\begin{equation}
\int \tilde {dk}k^\mu \hat I(k)=0,  \label{2.1}
\end{equation}

\begin{equation}
\int \tilde {dk}{\hat \phi }(k)\frac \partial {\partial P_\mu }F((P\cdot
k),P^2)=0,  \label{2.2}
\end{equation}
that is, we require the contribution to the collective variables from these
new fields to be zero. This is analogous to the definition of the relative
variables for the relativistic string model in the quoted reference \cite
{CLL} (see Eq.(2.20) of that paper).

A simple choice for ${\hat I}(k)$ and ${\hat\phi}(k)$ seems to be

\begin{equation}
\hat I(k)=I(k)-F((P\cdot k),P^2),  \label{2.3}
\end{equation}

\begin{equation}
{\hat \phi}(k) = \phi(k) - (k\cdot X).  \label{2.4}
\end{equation}

However, these variables are not canonical: they satisfy

\begin{equation}
\{\hat I(k),\ {\hat \phi }(k)\}=\Delta (k,k^{\prime }),  \label{2.5}
\end{equation}

\noindent where

\begin{equation}
\Delta (k,k^{\prime })=\Omega (k)\delta ^3(\vec k-\vec k^{\prime
})-k^{\prime \mu }\frac \partial {\partial P^\mu }F((P\cdot k),P^2),
\label{2.6}
\end{equation}

\noindent with the following properties:

\begin{equation}
\int\tilde{dk^{\prime}}\Delta(k, k^{\prime}) \Delta(k^{\prime},
k^{\prime\prime}) = \Delta(k, k^{\prime\prime}),  \label{2.7}
\end{equation}

\begin{equation}
\int\tilde{dk^{\prime}} k^{\mu}\Delta(k, k^{\prime}) = 0,  \label{2.8}
\end{equation}

\begin{equation}
\int \tilde {dk^{\prime }}\Delta (k,k^{\prime })\frac \partial {\partial
P^\mu }F((P\cdot k^{\prime }),P^2)=0.  \label{2.9}
\end{equation}

They have some Poisson brackets good for relative variables

\begin{equation}
\{P^\mu ,\ {\hat \phi }(k)\}=0,\quad \{P^\mu ,\ \hat I(k)\}=0,\quad \{X^\mu
,\ \hat I(k)\}=0;  \label{2.10}
\end{equation}

\noindent however

\begin{equation}
\{ X^{\mu},\ {\hat \phi}(k)\} \not= 0.  \label{2.11}
\end{equation}

So they are not canonical relative coordinates; nevertheless, we will go on
using them since, as can be ascertained with some algebra, they have the
interesting property of separating the generators $M_{\mu \nu }$ of the
Lorentz group in a collective part, depending solely on $X$ and $P$, and a
second part, depending on $\hat I(k)$ and $\hat \phi (k)$, as: 
\begin{equation}
M_{\mu \nu }=L_{\mu \nu }\left[ X,P\right] +S_{\mu \nu }\left[ \hat I,\hat 
\phi \right] .  \label{(M=L+S).1}
\end{equation}
In spite of Eq. (\ref{2.11}), due to the condition (\ref{2.1}), these two
parts are in involution, that is their Poisson bracket is zero: this means
that the part $S_{\mu \nu }\left[ \hat I,\hat \phi \right] $ of $M_{\mu \nu
} $ depending solely on subsidiary variables is what must be referred to as
the {\it relative part}.

\subsection{The relative variable ${\cal H}(k)$.}

Two new fields ${\cal H}(k)$ and ${\cal K}(k)$, going to play the role of
canonical relative variables, can be defined in order to fulfill the
integral constraints on $\hat I(k)$ and ${\hat \phi }(k)$: 
\begin{equation}
\begin{array}{cc}
\displaystyle \int 
\tilde {dk}k^\mu \hat I(k)=0, & 
\displaystyle \int 
\tilde {dk}{\hat \phi }(k)%
{\displaystyle {\partial \over \partial P_\mu }}
F((P\cdot k),P^2)=0.
\end{array}
\label{fred.1}
\end{equation}

Let us deal with the first constraint on $\hat I(k)$, obtained putting $\mu
=0$: 
\begin{equation}
\displaystyle \int 
d^3k\hat I(k)=0;  \label{div.1}
\end{equation}
if the quantity $\int d^3k\hat I(k)$ is zero, it can be thought of as
resulting from the integration of some divergence: 
\begin{equation}
\begin{array}{cc}
\hat I\left( k\right) =\vec \nabla \cdot \vec J\left( k\right) , & 
\lim\limits_{\left| \vec k\right| \rightarrow +\infty }\vec J\left( k\right)
=0.
\end{array}
\label{div.5}
\end{equation}

This relationship (\ref{div.5}) can now be used in the other 3 constraints
on $\hat I\left( k\right) $, for $\mu =i$: by replacing $\hat I\left(
k\right) $ with $\vec \nabla \cdot \vec J\left( k\right) $ inside the
condition 
\[
\displaystyle \int 
\tilde {dk}k^i\hat I(k)=0 
\]
one gets: 
\[
\displaystyle \int 
\frac{d^3k}{\omega \left( k\right) }k^i\partial _hJ^h(k)=0, 
\]
which can be re-written as 
\begin{equation}
\displaystyle \int 
d^3k\left[ \partial _h\left( \frac{k^i}{\omega \left( k\right) }%
J^h(k)\right) -J^h(k)\partial _h\left( \frac{k^i}{\omega \left( k\right) }%
\right) \right] =0.  \label{div.6}
\end{equation}
The integral of the first term is zero under the hypothesis (\ref{div.5}),
so that the constraints on $\hat I\left( k\right) $ with $\mu =i$ are
equivalent to: 
\begin{equation}
\displaystyle \int 
J^h(k)\partial _h\left( \frac{k^i}{\omega \left( k\right) }\right) d^3k=0.
\label{div.7}
\end{equation}

One can suppose that there exist some field ${\rm \tilde H}\left( k\right) $
so that 
\begin{equation}
-J^h(k)\partial _h\left( \frac{k^i}{\omega \left( k\right) }\right)
=\partial _i{\rm \tilde H}\left( k\right) ;  \label{div.8}
\end{equation}
in this way a suitable ultraviolet behaviour of ${\rm \tilde H}\left(
k\right) $ allows (\ref{div.7}) to take place. By performing the
calculations indicated in (\ref{div.8}), making the scalar product with $%
\vec k$ of both r.h.s. and l.h.s., exploiting again the definition (\ref
{div.5}) and the expression of $\omega \left( k\right) $ in terms of $\vec k$%
, one finally gets: 
\begin{equation}
\hat I\left( k\right) ={\rm \tilde D\tilde H}\left( k\right) ,  \label{div.9}
\end{equation}
with 
\begin{equation}
{\rm \tilde D}=-\frac{\omega \left( k\right) }{m^2}\left[ m^2\vec \nabla
\cdot \vec \nabla +4\left( \vec k\cdot \vec \nabla \right) +\left( \vec k%
\cdot \vec \nabla \right) ^2\right] .  \label{div.10}
\end{equation}
If the new field ${\rm \tilde H}\left( k\right) $ defined as in (\ref{div.9}%
) is introduced, the subsidiary variable $\hat I\left( k\right) $
immediately fulfills the constraints (\ref{fred.1}).

There is one problem with using this ${\rm \tilde H}\left( k\right) $ as a
relative variable: it is not scalar under Poincar\'e transformations. In
fact, the differential operator ${\rm \tilde D}$ does not behave like a
scalar, while $\hat I\left( k\right) $ does (so that from Eq.(\ref{div.9})
the non-scalar behaviour of ${\rm \tilde H}\left( k\right) $ is evident).

This is less catastrophic than one could think, indeed it is easily shown
that 
\begin{equation}
\begin{array}{cc}
{\rm \tilde D}=-%
{\displaystyle {1 \over m^2}}
\left( m^2\Delta -3\right) \omega \left( k\right) , & \Delta =\vec \nabla
\cdot \vec \nabla +%
{\displaystyle {2 \over m^2}}
\left( \vec k\cdot \vec \nabla \right) +%
{\displaystyle {1 \over m^2}}
\left( \vec k\cdot \vec \nabla \right) ^2.
\end{array}
\label{2.20}
\end{equation}
The new operator $\Delta $ is scalar under Poincar\'e transformations, and
has a clear geometrical meaning, as we shall show few lines below.

Due to Eq.(\ref{2.20}) one can introduce one new scalar differential
operator $D$ and one new scalar field ${\cal H}$ 
\begin{equation}
\begin{array}{cc}
D=3-m^2\Delta , & {\cal H}\left( k\right) =%
{\displaystyle {\omega \left( k\right)  \over m^2}}
{\rm \tilde H}\left( k\right)
\end{array}
\label{div.12}
\end{equation}
such that 
\begin{equation}
\hat I\left( k\right) =D{\cal H}\left( k\right)  \label{2.26}
\end{equation}
identically gives rise to the constraints on $\hat I\left( k\right) $ as in
Eqs.(\ref{fred.1}). See Appendix C for the study of the operator $D$.

The Poincar\'e-scalar nature of the operator $D$ is clear: in fact $D$ is
naturally defined in terms of {\it the Laplace-Beltrami operator on the mass
shell submanifold} \cite{bms-matelonghi} (referred to as $H_3^1$ and studied
in \cite{RLN}), that has equation 
\begin{equation}
k^\mu k_\mu =m^2,\;{\rm and\;}k^0>0.  \label{2.26.bis}
\end{equation}
This Laplace-Beltrami operator on $H_3^1$ turns out to be that $\Delta $
introduced in Eq.(\ref{2.20}).

$\Delta $ is the only invariant differential operator of the second order on 
$H_3^1$ (no invariant first order differential operator exists), and it is
defined as

\begin{equation}
\Delta =-{\frac 1{\sqrt{\hat \eta }}}{\frac \partial {\partial k^i}}{\hat 
\eta }^{ij}\sqrt{{\hat \eta }}{\frac \partial {\partial k^j}}  \label{2.12}
\end{equation}
in terms of $H_3^1$ geometry, where ${\hat \eta }=\det \parallel {\hat \eta }%
_{ij}\parallel $ and where ${\hat \eta }_{ij}$ is the embedding-induced
metric on the mass shell submanifold \cite{bms-matelonghi}, \cite{RLN}.
Explicitly, it has the form given in (\ref{2.20}).

$\Delta $ is {\it invariant} under a Lorentz transformation $\Lambda $, that
is represented as

\begin{equation}
k^i\rightarrow k^{\prime \ i}=\Lambda _j^ik^j+\Lambda _0^i\omega (k),
\label{2.21}
\end{equation}
on $\vec k\in H_3^1$, and is {\it formally self-adjoint} with respect to the
invariant measure $\tilde {dk}$. It is an {\it elliptic} operator, and has
the property

\begin{equation}
\Delta k^\mu ={\frac 3{m^2}}k^\mu .  \label{2.24}
\end{equation}

Then the differential operator $D$

\begin{equation}
D=3-m^2\Delta  \label{2.25}
\end{equation}
is formally self-adjoint too with respect to the scalar product of $L_2(%
\tilde {dk})$, and Eq.(\ref{2.24}) suggests that the four functions 
\begin{equation}
\begin{array}{ccc}
k^\mu \left( k\right) & : & \left\{ 
\begin{array}{l}
k^0\left( k\right) =\omega \left( k\right) , \\ 
k^i\left( k\right) =k^i
\end{array}
\right.
\end{array}
\label{div.14}
\end{equation}
are all among its null modes. Thus, assuming (\ref{2.26}) it is possible to
get

\begin{equation}
\int \tilde {dk}k^\mu \hat I(k)=\int \tilde {dk}k^\mu D{\cal H}(k)=\int 
\tilde {dk}{\cal H}(k)Dk^\mu =0  \label{2.12''}
\end{equation}
identically (self-adjointness of $D$ cancels any boundary term in (\ref
{2.12''}) by definition).

However, the assumption (\ref{2.26}) is not in general consistent. Its
consistency rests upon a set of conditions that $\hat I(k)$ must satisfy,
which will be thoroughly discussed in Section IV. Let us for the present 
{\it assume} Eq.(\ref{2.26}) and develop its consequences:\ we postpone the
discussion of its mathematical consistency just for simplicity, since here
we are more interested in the physical picture than in even important
mathematical conditions.

After this construction the field $I(k)$ can be expressed as a non linear
function of $P^\mu $ and as a linear function of the field ${\cal H}(k)$,
that is

\begin{equation}
I(k)=D{\cal H}(k)+F((P\cdot k),P^2),  \label{5.91}
\end{equation}
while the inverse transformation reads:

\begin{equation}
\begin{array}{l}
P^\mu =%
\displaystyle \int 
\tilde {dk}k^\mu I(k), \\ 
{\cal H}(k)=%
\displaystyle \int 
\tilde {dk^{\prime }}G(k,k^{\prime })\left[ I(k^{\prime })-F((P\cdot
k^{\prime }),P^2)\right]
\end{array}
\label{5.101}
\end{equation}
(where in the second equation $P^\mu $ must be considered a functional of $%
I(k)$ and where $G(k,k^{\prime })$ is the Green function of the operator $D$%
, which is discussed in the Appendix D).

Let us finally show that the new field ${\cal H}(k)$ is a relative variable,
since we can easily check that its Poisson brackets with $X$ and $P$ are
zero: 
\begin{equation}
\begin{array}{cc}
\left\{ {\cal H}(k),P^\mu \right\} =0, & \left\{ {\cal H}(k),X^\mu \right\}
=0.
\end{array}
\label{h.relativo}
\end{equation}
In fact, since

\begin{equation}
\{\phi(k^{\prime}),\ \hat{I}(k^{\prime\prime})\} = -
\Delta(k^{\prime\prime}, k^{\prime}),  \label{5.19}
\end{equation}

\noindent we get

\begin{equation}
\{X^\mu ,\ {\cal H}(k)\}=-\int \tilde {dk}^{\prime }\int \tilde {dk}^{\prime
\prime }G(k,k^{\prime \prime }){\frac{\partial F((P\cdot k^{\prime }),P^2)}{%
\partial P_\mu }}\Delta (k^{\prime \prime },k^{\prime })=0,  \label{5.20}
\end{equation}

\noindent from Eq.(\ref{2.9}). Besides, ${\cal H}(k)$ and $P^\mu $ have zero
Poisson brackets since they are both functionally dependent from the same
field $I(k)$ only.

\subsection{The second relative variable ${\cal K}(k)$.}

Here the canonical variables conjugated to the relative field ${\cal H}%
\left( k\right) $ are introduced. We have already defined the field $\hat 
\phi (k)$ in Eq.(\ref{2.4}) as a preliminary choice for the relative phases

\begin{equation}
\hat \phi (k)=\phi (k)-(k\cdot X);  \label{5.111}
\end{equation}
now we start from the fact that the subsidiary variable $\hat \phi (k)$
satisfies the integral constraint: 
\begin{equation}
\int \tilde {dk}{\hat \phi }(k)\frac \partial {\partial P_\mu }F((P\cdot
k),P^2)=0;  \label{vincolo.phi}
\end{equation}

We'll define some relative field ${\cal K}\left( k\right) $ canonically
conjugated to ${\cal H}\left( k\right) $ so that (\ref{vincolo.phi}) is an
identity when $\hat \phi =\hat \phi \left[ {\cal K},...\right] $.

We choose to relate the new variable to the old one in a linear way 
\begin{equation}
\hat \phi (k)=\int d\tilde k^{\prime }A\left( k^{\prime },k\right) {\cal K}%
(k^{\prime }),  \label{phi.1}
\end{equation}
so that matching (\ref{phi.1}) with (\ref{vincolo.phi}) one gets: 
\begin{equation}
\int d\tilde k^{\prime }{\cal K}(k^{\prime })\int \tilde {dk}A\left(
k^{\prime },k\right) \frac \partial {\partial P_\mu }F((P\cdot
k),P^2)=0,\;\forall \;{\cal K}.  \label{phi.2}
\end{equation}
In order to have a null integral for every possible form of the function $%
{\cal K}$ one has to admit 
\begin{equation}
\int \tilde {dk}A\left( k^{\prime },k\right) \frac \partial {\partial P_\mu }%
F((P\cdot k),P^2)=0,  \label{phi.3}
\end{equation}
and since we've already noted that 
\[
\displaystyle \int 
\tilde {dk}\Delta \left( k^{\prime \prime },k\right) 
{\displaystyle {\partial \over \partial P_\mu }}
F((P\cdot k),P^2)=0, 
\]
one can always get a solution for (\ref{phi.3}) with the position: 
\begin{equation}
A\left( k^{\prime },k\right) =%
\displaystyle \int 
d\tilde k^{\prime \prime }R\left( k^{\prime },k^{\prime \prime }\right)
\Delta \left( k^{\prime \prime },k\right) .  \label{phi.4}
\end{equation}

The relative variable ${\cal K}(k)$ has been constructed so that 
\begin{equation}
\hat \phi (k)=\int d\tilde k^{\prime }%
\displaystyle \int 
d\tilde k^{\prime \prime }{\cal K}(k^{\prime })R\left( k^{\prime },k^{\prime
\prime }\right) \Delta \left( k^{\prime \prime },k\right) ,  \label{phi.5}
\end{equation}
and the distribution $R\left( k^{\prime },k^{\prime \prime }\right) $ can be
identified by requiring the field $\hat \phi (k)$ to have the right Poisson
bracket with $\hat I\left( k\right) $; it is very easy to check that: 
\[
\begin{array}{ccc}
\left\{ \hat I\left( k\right) ,\hat \phi \left( k^{\prime }\right) \right\}
=\Delta \left( k,k^{\prime }\right) & \Rightarrow & D_{k^{\prime }}R\left(
k^{\prime },k^{\prime \prime }\right) =\Omega \left( k^{\prime }\right)
\delta ^3\left( \vec k^{\prime }-\vec k^{\prime \prime }\right) .
\end{array}
\]
The conclusion is that {\it the coefficients }$R\left( k^{\prime },k^{\prime
\prime }\right) ${\it \ must be a Green function for our operator }$D$, so
that we can choose to write: 
\begin{equation}
\hat \phi (k)=\int d\tilde k^{\prime }%
\displaystyle \int 
d\tilde k^{\prime \prime }{\cal K}(k^{\prime })G\left( k^{\prime },k^{\prime
\prime }\right) \Delta \left( k^{\prime \prime },k\right) .
\label{phi.da.K.1}
\end{equation}

It is easily found that {\it two} different inversions of (\ref{phi.da.K.1})
are possible

\begin{equation}
\begin{array}{cc}
{\cal K}(k)=D\hat \phi (k), & {\cal K}(k)=D\phi (k):
\end{array}
\label{5.1311}
\end{equation}
this fact suggests that the transformation $\phi \rightarrow {\cal K}$ is
some sort of projection, which will be seen to be equivalent to a limiting
assumption on the form of $\phi \left( k\right) $ in order for the
transformation $\Phi ,\Pi \rightarrow X,P,{\cal H},{\cal K}$ to be
consistent.

Eqs.(\ref{5.1311}) give us concrete information about the ultraviolet
behaviour of ${\cal K}(k)$:

\begin{equation}
|{\cal K}(k)|\simeq \left| \vec k\right| ^{1-\epsilon },\quad {\rm {for}%
\quad {any}\quad \epsilon >0,}  \label{5.132}
\end{equation}
since any linear dependence of $k^\mu $ in $\phi \left( k\right) $ is
annihilated by $D$.

The two fields ${\cal H}$ and ${\cal K}\ $are canonically conjugated, as we
may easily verify by observing that the Green function $G$ is symmetric in
its arguments. So one has:

\begin{equation}
\{{\cal H}(k),\ {\cal K}(k^{\prime })\}=\Omega (k)\delta ^3\left( \vec k-%
\vec k^{\prime }\right) .  \label{5.15}
\end{equation}

Moreover, the Poisson bracket of ${\cal H}(k)$ with $P^{\mu}$ and $X^{\mu} $
is zero. Indeed

\begin{eqnarray}
\{ P^{\mu},\ {\cal K}(k)\} = D\{ P^{\mu},\ \phi(k)\} =  \nonumber \\
= D\int\tilde{dk}^{\prime} {k^{\prime}}^{\mu} \{ I(k^{\prime}),\ \phi(k)\} =
D k^{\mu} = 0,  \label{5.16}
\end{eqnarray}

\noindent and

\begin{equation}
\begin{array}{l}
\{X^\mu ,\ {\cal K}(k)\}=D_k%
\displaystyle \int 
d\tilde k^{\prime }\phi (k^{\prime })\left\{ 
{\displaystyle {\partial F((P\cdot k^{\prime }),P^2) \over \partial P_\mu }}
,\ \phi (k)\right\} = \\ 
\\ 
=D_k%
\displaystyle \int 
d\tilde k^{\prime }\phi (k^{\prime }) 
{\displaystyle {\partial ^2F((P\cdot k^{\prime }),P^2) \over {\partial P_\mu }{\partial P_\nu }}}
\{P_\nu ,\ \phi (k)\}= \\ 
\\ 
=\left[ 
\displaystyle \int 
d\tilde k^{\prime }\phi (k^{\prime }) 
{\displaystyle {\partial ^2F((P\cdot k^{\prime }),P^2) \over {\partial P_\mu }{\partial P_\nu }}}
\right] D_k\ k_\nu =0,
\end{array}
\label{5.17}
\end{equation}
since $k_\nu \in \ker D$.

As we shall stress discussing the consistency conditions of the whole
transformation, Eqs.(\ref{5.1311}) should be assumed as {\it the definition}
of ${\cal K}(k)$; thus, it would be more ''mathematically correct'' to start
from the first of (\ref{5.1311}) and show that with the inversion (\ref
{phi.da.K.1}) the variable $\hat \phi (k)$ has the nice properties it has to
have. The choice of the present exposition simply underlines the physical
derivation of ${\hat \phi =\hat \phi }\left[ {\cal K},...\right] $ directly
from the relations that must hold for $\hat \phi (k)$.

\subsection{Field variables and Lorentz algebra in terms of the new
variables.}

The new canonical set has the following nonzero Poisson brackets

\begin{equation}
\{ P^{\mu},\ X^{\nu}\} = \eta^{\mu \nu},  \label{5.21}
\end{equation}

\begin{equation}
\{{\cal H}(k),\ {\cal K}(k^{\prime })\}=\Omega (k)\delta ^3\left( \vec k-%
\vec k^{\prime }\right) ,  \label{5.22}
\end{equation}

\noindent and the other possible combinations all vanish.

These variables, except $X^0$, are constants of motion, since they have zero
Poisson bracket with the Hamiltonian $P^0$. Thus the canonical basis we were
looking for $\{P^\mu ,X^\mu ,{\cal H}(k),{\cal K}(k)\}$ is some set of
Hamilton-Jacobi data.

The canonical transformation from $\left\{ I\left( k\right) ,\phi \left(
k\right) \right\} $ is

\begin{equation}
P^{\mu} = \int\tilde{dk} k^{\mu} I(k),  \label{5.23}
\end{equation}

\begin{equation}
X^\mu =\int \tilde {dk}\frac{\partial F((P\cdot k),P^2)}{\partial P_\mu }%
\phi (k),  \label{5.24}
\end{equation}

\begin{equation}
{\cal H}(k) = \int\tilde{dk}^{\prime} G(k, k^{\prime})\left[ I(k^{\prime}) -
F((P\cdot k^{\prime}), P^2)\right],  \label{5.25}
\end{equation}

\begin{equation}
{\cal K}(k) = D\phi(k),  \label{5.255}
\end{equation}

\noindent and the inverse reads:

\begin{equation}
I(k)=D{\cal H}(k)+F((P\cdot k),P^2),  \label{5.26}
\end{equation}

\begin{equation}
\phi (k)=(k\cdot X)+\int d\tilde k^{\prime }%
\displaystyle \int 
d\tilde k^{\prime \prime }{\cal K}(k^{\prime })G\left( k^{\prime },k^{\prime
\prime }\right) \Delta \left( k^{\prime \prime },k\right)  \label{a5.27}
\end{equation}

We may give the original field variables in terms of the new variables as:

\begin{equation}
\begin{array}{l}
\Phi (x)= \\ 
\\ 
=%
\displaystyle \int 
\tilde {dk}\sqrt{D{\cal H}(k)+F((P\cdot k),P^2)}\left[ e^{i(k\cdot
(X-x))+i\int d\tilde k^{\prime }\int d\tilde k^{\prime \prime }{\cal K}%
(k^{\prime })G\left( k^{\prime },k^{\prime \prime }\right) \Delta \left(
k^{\prime \prime },k\right) }+c.c\right] ,
\end{array}
\label{oldfield.1}
\end{equation}

\begin{equation}
\begin{array}{l}
\Pi (x)= \\ 
\\ 
=-i%
\displaystyle \int 
\tilde {dk}\omega (k)\sqrt{D{\cal H}(k)+F((P\cdot k),P^2)}\left[ e^{i(k\cdot
(X-x))+i\int d\tilde k^{\prime }\int d\tilde k^{\prime \prime }{\cal K}%
(k^{\prime })G\left( k^{\prime },k^{\prime \prime }\right) \Delta \left(
k^{\prime \prime },k\right) }-c.c\right] .
\end{array}
\label{oldfield.2}
\end{equation}

It's particularly interesting to study the Poincar\'e algebra in terms of
the new variables: the generators of the group naturally decompose into
collective and relative parts, as much as in the case of a system of
particles, so our main goal has been achieved.

The Eq.(\ref{1.19}), (\ref{1.20}), and (\ref{1.21}) give the generators of
the Lorentz algebra in terms of the modulus-phase variables. We want to
stress that, in terms of the new collective and relative variables, we get
for the generators of the Lorentz group

\begin{equation}
M_{ij}=L_{ij}\left[ X,P\right] +S_{ij}\left[ {\cal H},{\cal K}\right] ,
\label{6.1}
\end{equation}

\noindent where

\begin{equation}
L_{ij}\left[ X,P\right] =X^iP^j-X^jP^i,  \label{6.2}
\end{equation}

\begin{equation}
S_{ij}\left[ {\cal H},{\cal K}\right] =\int \tilde {dk}{\cal H}(k)\left( k^i{%
\frac \partial {\partial k^j}}-k^j{\frac \partial {\partial k^i}}\right) 
{\cal K}(k).  \label{6.3}
\end{equation}

Moreover, each generator $L_{ij}\left[ X,P\right] \ $and $S_{ij}\left[ {\cal %
H},{\cal K}\right] \ $separately satisfies the Lorentz algebra, given by Eq.(%
\ref{A.11}). The calculus is a boring one, and is sketched in Appendix E.

Analogously we get

\begin{equation}
M_{0j}=L_{0j}\left[ X,P\right] +S_{0j}\left[ {\cal H},{\cal K}\right] ,
\label{6.4}
\end{equation}

\noindent where

\begin{equation}
L_{0j}\left[ X,P\right] =\left( x_0+X_0\right) P_j-X_jP_0,  \label{6.5}
\end{equation}

\begin{equation}
S_{0j}\left[ {\cal H},{\cal K}\right] =-\int \tilde {dk}{\cal H}(k)\omega (k)%
{\frac \partial {\partial k^j}}{\cal K}(k).  \label{6.6}
\end{equation}

Putting shortly 
\[
\begin{array}{cc}
L_{\mu \nu }\left[ X,P\right] =M_{\mu \nu }^{\left( 0\right) }, & S_{\mu \nu
}\left[ {\cal H},{\cal K}\right] =M_{\mu \nu }^{\left( 1\right) },
\end{array}
\]
their algebra reads

\begin{equation}
\left\{ M_{ij}^{(r)},\ M_{hk}^{(s)}\right\} =\left[ \delta
_{ih}M_{jk}^{(r)}+\delta _{jk}M_{ih}^{(r)}-\delta _{ik}M_{jh}^{(r)}-\delta
_{jh}M_{ik}^{(r)}\right] \delta _{\left( r\right) \left( s\right) },
\label{6.7}
\end{equation}

\begin{equation}
\left\{ M_{ij}^{(r)},\ M_{0k}^{(s)}\right\} =\left[ \delta
_{ik}M_{0j}^{(r)}-\delta _{jk}M_{0i}^{(r)}\right] \delta _{\left( r\right)
\left( s\right) },  \label{6.8}
\end{equation}

\begin{equation}
\left\{ M_{0i}^{(r)},\ M_{0j}^{(s)}\right\} =-M_{ij}^{(r)}\delta _{\left(
r\right) \left( s\right) },  \label{6.9}
\end{equation}

\noindent where $\left( r\right) ,\left( s\right) =0,1$.

The canonical decomposition of the angular momenta in a collective $L_{\mu
\nu }\left[ X,P\right] $ part and a relative one $S_{\mu \nu }\left[ {\cal H}%
,{\cal K}\right] $ is achieved in this way.

\section{The consistency conditions.}

Now it is necessary to come to the consistency conditions of the whole
mechanism elaborated in order to perform the transformation 
\begin{equation}
\begin{array}{ccc}
\Phi \left( x\right) ,\Pi \left( x\right) & \mapsto & X^\mu ,P^\mu ;{\cal H}%
(k),{\cal K}(k)
\end{array}
\label{quattro.1}
\end{equation}
for the Klein-Gordon field; the key-equations of our approach are the
conditions 
\begin{equation}
\begin{array}{cc}
\displaystyle \int 
d\tilde kk^\mu \hat I\left( k\right) =0, & 
\displaystyle \int 
d\tilde k%
{\displaystyle {\partial F\left( P,k\right)  \over \partial P^\mu }}
\hat \phi \left( k\right) =0
\end{array}
\label{quattro.2}
\end{equation}
on the subsidiary variables, that have been identically fulfilled by
putting: 
\begin{equation}
\left\{ 
\begin{array}{l}
I\left( k\right) =D{\cal H}(k)+F\left( P,k\right) =\hat I\left( k\right)
+F\left( P,k\right) , \\ 
\phi \left( k\right) =\left( k\cdot X\right) +%
\displaystyle \int 
d\tilde k^{\prime }%
\displaystyle \int 
d\tilde k^{\prime \prime }{\cal K}\left( k^{\prime }\right) G\left(
k^{\prime },k^{\prime \prime }\right) \Delta \left( k^{\prime \prime
},k\right) .
\end{array}
\right.  \label{quattro.3}
\end{equation}

One has to identify carefully those functional spaces for $\Phi $ and $\Pi $
within which these Eqs.(\ref{quattro.3}) are consistent.

The first of Eqs.(\ref{quattro.3}) is not in general consistent: we must
check the effect that the zero modes of the operator $D$ may have on it. The
analysis of Appendix C, where we study the null space of $D$, shows that
this space has no nontrivial intersection with the $L_2\left( d\tilde k%
\right) $ space. A na\"\i ve application of the Fredholm alternative theorem
could imply that the operator $D$ is a 1-1 application within $L_2\left( d%
\tilde k\right) $. However {\it the usual form of Fredholm theorem applies
only to operators which are compact or have compact inverse}: if this is the
case, the theorem asserts that an operator with empty null space in $L_2$ is
1-1, and that an equation like the first one in (\ref{quattro.3}) is
consistent.

Unfortunately {\it our operator }$D${\it \ is not compact, nor its inverse}
as determined by the Green function studied in Appendix D, and this all can
be easily checked by observing that the inverse operator, whose kernel is
the Green function, does not change the leading asymptotic term of a
function of $\vec k$ as $\left| \vec k\right| \rightarrow \infty $.

Let us start with checking the consistency conditions of the first of (\ref
{quattro.3}): 
\begin{equation}
\hat I(k)=D{\cal H}(k).  \label{fred.2.a}
\end{equation}

If $f_0$ is a function belonging to the kernel of $D$%
\begin{equation}
f_0\in \ker D  \label{fred.3}
\end{equation}
and has suitable properties so that the integral 
\[
\displaystyle \int 
d\tilde k\hat I(k)f_0\left( k\right) 
\]
is a finite quantity, one can see that an unavoidable consequence of (\ref
{fred.2.a}) is: 
\begin{equation}
\displaystyle \int 
d\tilde k\hat I(k)f_0\left( k\right) =%
\displaystyle \int 
d\tilde kf_0\left( k\right) D{\cal H}\left( k\right) .  \label{fred.3.bis}
\end{equation}
Let us define 
\begin{equation}
T_\partial \left[ f,g\right] =%
\displaystyle \int 
g\left( k\right) Df\left( k\right) d\tilde k-%
\displaystyle \int 
f\left( k\right) Dg\left( k\right) d\tilde k  \label{fred.3.a}
\end{equation}
and then re-express the r.h.s. of Eq.(\ref{fred.3.bis}) as: 
\begin{equation}
\displaystyle \int 
d\tilde kf_0\left( k\right) D{\cal H}\left( k\right) =%
\displaystyle \int 
d\tilde k{\cal H}\left( k\right) Df_0\left( k\right) +T_\partial \left[ 
{\cal H},f_0\right] =T_\partial \left[ {\cal H},f_0\right]
\label{fred.3.a.bis}
\end{equation}
from Eq.(\ref{fred.3}). Assuming the behaviour of ${\cal H}$ so that the
boundary term $T_\partial \left[ {\cal H},f_0\right] $ vanishes 
\begin{equation}
T_\partial \left[ {\cal H},f_0\right] =0  \label{fred.3.b}
\end{equation}
we have: 
\begin{equation}
\displaystyle \int 
d\tilde k\hat I\left( k\right) f_0\left( k\right) =0,  \label{fred.4}
\end{equation}
that must hold as an integrability condition.

The condition (\ref{fred.3.b}) can be realized for those zeroes $f_0$ of $D$
that render the integral $\int d\tilde kIf_0$ a finite quantity with a
suitable assumption on ${\cal H}\left( k\right) $: those zeroes are the only
ones for which the whole discussion makes sense. The surface term $%
T_\partial \left[ f_0,{\cal H}\right] $ is given by 
\[
\begin{array}{l}
T_\partial \left[ f_0,{\cal H}\right] =%
\displaystyle \int 
\left[ f_0\left( k\right) D{\cal H}\left( k\right) -{\cal H}\left( k\right)
Df_0\left( k\right) \right] d\tilde k \\ 
\\ 
=-%
{\displaystyle {1 \over 2\left( 2\pi \right) ^3}}
\displaystyle \int 
d^3k%
{\displaystyle {\partial \over \partial k^i}}
\left[ 
{\displaystyle {m^2\delta ^{ij}+k^ik^j \over \omega \left( k\right) }}
\right] \left[ f_0\left( k\right) 
{\displaystyle {\partial \over \partial k^j}}
{\cal H}\left( k\right) -{\cal H}\left( k\right) 
{\displaystyle {\partial \over \partial k^j}}
f_0\left( k\right) \right] ,
\end{array}
\]
or, putting $\vec x=m^{-1}\vec k$ and $r=\left| \vec x\right| $, 
\begin{equation}
\begin{array}{l}
T_\partial \left[ f_0,{\cal H}\right] = \\ 
\\ 
=-%
{\displaystyle {m^2 \over 2\left( 2\pi \right) ^3}}
\left\{ 
\mathop{\rm lim}
\limits_{r\rightarrow \infty }-%
\mathop{\rm lim}
\limits_{r\rightarrow 0}\right\} 
\displaystyle \int 
d^2\Omega r^2\sqrt{1+r^2}\left[ f_0\left( x\right) 
{\displaystyle {\partial \over \partial r}}
{\cal H}\left( x\right) -{\cal H}\left( x\right) 
{\displaystyle {\partial \over \partial r}}
f_0\left( x\right) \right] .
\end{array}
\label{fred.3.ter}
\end{equation}

In order to estimate it, we may assume for ${\cal H}$ the following
asymptotic behaviours: 
\[
\left\{ 
\begin{array}{l}
\begin{array}{cccc}
{\cal H}\left( x\right) \sim \alpha \left( \hat x\right) r^{-1+\varepsilon }
& {\rm as} & r\rightarrow 0, & \varepsilon >0,
\end{array}
\\ 
\begin{array}{cccc}
{\cal H}\left( x\right) \sim \beta \left( \hat x\right) r^{-3-\sigma } & 
{\rm as} & r\rightarrow \infty , & \sigma >0,
\end{array}
\end{array}
\right. 
\]
where $\alpha \left( \hat x\right) $ and $\beta \left( \hat x\right) $ are
unknown functions of the direction of $\vec x$ only.

This asymptotic behaviour of ${\cal H}$ is a consequence of the assumptions
on $I\left( k\right) $, i.e. the existence of the generators of the
Poincar\'e group in Eqs.(\ref{1.19}), (\ref{1.20}) and (\ref{1.21}). The
addition of any null mode of $D$ to ${\cal H}$ has no effect on the surface
term.

In Appendix C we determine the following behaviour of the null modes of $D$,
see (\ref{B.3}), (\ref{B.9}) and (\ref{B.91}): 
\[
\left\{ 
\begin{array}{l}
v_{1,-3,l}^{\left( 0\right) }\left( r,\vartheta ,\varphi \right) \sim r^l,
\\ 
\begin{array}{ccc}
v_{2,-3,l}^{\left( 0\right) }\left( r,\vartheta ,\varphi \right) \sim
r^{-l-1}, & \forall & \varphi ,\vartheta
\end{array}
\end{array}
\right. 
\]
as $r\rightarrow 0$, and 
\[
\left\{ 
\begin{array}{l}
v_{1,-3,l}^{\left( 0\right) }\left( r,\vartheta ,\varphi \right) \sim a_lr,
\\ 
\begin{array}{cccc}
v_{2,-3,l=0,1}^{\left( 0\right) }\left( r,\vartheta ,\varphi \right) \sim
b_lr, & v_{2,-3,l>1}^{\left( 0\right) }\left( r,\vartheta ,\varphi \right)
\sim b_lr^{-3}, & \forall & \varphi ,\vartheta
\end{array}
\end{array}
\right. 
\]
as $r\rightarrow \infty $, where $a_l$ and $b_l$ are constants (being the $v$%
's those null modes of $D$ as defined in (\ref{B.91.a}), (\ref{B.91.b}), (%
\ref{B.91.c}) and (\ref{B.91.d})). It is easily seen that $T_\partial \left[
v_{1,-3,l,m}^{\left( 0\right) },{\cal H}\right] $ is always zero. For $%
T_\partial \left[ v_{2,-3,l,m}^{\left( 0\right) },{\cal H}\right] $ we have
the following situation: for $\epsilon >l+1$ it is zero; it is finite and
different from zero if $\epsilon =l+1$, and is divergent if $\epsilon <l+1$.

Using as null modes of $D$ the functions $v_{1,-3,l,m}^{\left( 0\right)
}\left( k\right) $ and $v_{2,-3,l,m}^{\left( 0\right) }\left( k\right) $ in
the integrability condition (\ref{fred.4}) one sees that the first equation (%
\ref{quattro.3}) is consistent under the following condition for $\hat I%
\left( k\right) $: 
\begin{equation}
\hat P_{l,m}=%
\displaystyle \int 
d\tilde k\hat I\left( k\right) v_{1,-3,l,m}^{\left( 0\right) }\left(
k\right) =0  \label{X.1}
\end{equation}
and 
\begin{equation}
\hat Q_{l,m}=%
\displaystyle \int 
d\tilde k\hat I\left( k\right) v_{2,-3,l,m}^{\left( 0\right) }\left(
k\right) =0  \label{X.2}
\end{equation}
if $\hat I\left( k\right) $ behaves like $k^{-3+\epsilon }$, for $\epsilon
>l+1$, about the origin.

It is very important to observe that conditions (\ref{X.2}) are absent for $%
0<\epsilon \leq 1$, since in this case the integrals {\it diverge}, and the
procedure worked out from Eq.(\ref{fred.2.a}) to Eq.(\ref{fred.3.a.bis}) has
no meaning. Instead, both the l.h.s. of this last equation and the surface
term are divergent.

The conditions (\ref{X.1}) and (\ref{X.2}) make the transformation to
collective and relative variables consistent. They determine the quantities 
\begin{equation}
\begin{array}{cc}
P_{l,m}=%
\displaystyle \int 
d\tilde kI\left( k\right) v_{1,-3,l,m}^{\left( 0\right) }\left( k\right) , & 
Q_{l,m}=%
\displaystyle \int 
d\tilde kI\left( k\right) v_{2,-3,l,m}^{\left( 0\right) }\left( k\right)
\end{array}
\label{X.2.1}
\end{equation}
in terms of $P$, like: 
\begin{equation}
\begin{array}{cc}
P_{l,m}=%
\displaystyle \int 
d\tilde kF\left( P,k\right) v_{1,-3,l,m}^{\left( 0\right) }\left( k\right) ,
& Q_{l,m}=%
\displaystyle \int 
d\tilde kF\left( P,k\right) v_{2,-3,l,m}^{\left( 0\right) }\left( k\right).
\end{array}
\label{X.2.3}
\end{equation}
The first of Eqs.(\ref{X.2.3}) is an identity for $l=0,1$, see
Eq.(\ref{1.8}). Apart from this case these equations show that $P_{l,m}$'s and
$Q_{l,m}$'s are functionals of $P_{\mu}$, and exclude the possibility of using
them as other independent canonical degrees of freedom.

In particular the condition given by Eq.(\ref{X.1}) can be referred to as 
{\it no-supertranslation condition}: in fact it renders impossible the use
of the $P_{l,m}$'s (which are referred to as ''supertranslations'' in \cite
{bms-matelonghi}) as canonical variables independent with respect to $P$.

The set of conditions (\ref{X.1}) and (\ref{X.2}) must be understood as
restrictions on the configuration space of the field, and they single out
those configurations for which the transformations (\ref{quattro.1}) are
possible. The problem is that they are not so easily understandable in terms
of restrictions directly settled on $\Phi \left( \vec x,t\right) $ and $\Pi
\left( \vec x,t\right) $, since they correspond to very complicated
constraints on the fields.

Note however that when $\epsilon \in \left( 0,1\right] $ only conditions (%
\ref{X.1}) make sense, so it's possible to choose suitably $I\left( k\right) 
$ identifying the consistency conditions simply with no supertranslation
condition.

We can finally state that the transformation (\ref{fred.2.a}) from $I\left(
k\right) $ to ${\cal H}\left( k\right) $ makes sense for a well defined set
of field configurations, divided into two subsets. First of all, Eq.(\ref
{fred.2.a}) does not make sense if neither conditions (\ref{X.1}) nor (\ref
{X.2}) are fulfilled. It makes sense for those configurations with 
\begin{equation}
\hat I\left( k\right) \approx k^{-3+\epsilon },\ {\rm for\;}\epsilon \in
\left( 0,1\right] ,\ \left| \vec k\right| \simeq 0  \label{1st.fam}
\end{equation}
if they fulfill conditions (\ref{X.1}), and for those with 
\begin{equation}
\hat I\left( k\right) \approx k^{-3+\epsilon },\ {\rm for\;}\epsilon >l+1,\
\left| \vec k\right| \simeq 0  \label{2nd.fam}
\end{equation}
if they fulfill Eq.(\ref{X.1}) as well as Eq.(\ref{X.2}).

The condition (\ref{X.1}) excludes supertranslation generators $P_{l,m}$'s
from the context of independent canonical variables in order for the
definition of these ${\cal H}\left( k\right) $ to be consistent: we
already know that it is possible to give a symplectic realization of the BMS
algebra in terms of the Klein-Gordon phase space \cite{bms-matelonghi}, but the
existence of constraints on them, given by the consistency conditions necessary
for the
definition of canonical relative variables, is a very new fact. The problem
of finding whether no supertranslations condition is a fundamental need of
any collective-relative splitting framework, or it only affects the
particular path we have chosen, deserves much more study.

An analogous discussion can be presented for the variable $\phi \left(
k\right) $; however the situation is different, since now the relative
variable ${\cal K}\left( k\right) $ is {\it defined} by Eq. (\ref{5.1311});
moreover an $L_2$-like scalar product of $\phi \left( k\right) $ with the
null modes of $D$ does not exist, due to the linear behaviour of $\phi
\left( k\right) $ at the infinity; there can be a doubt about the validity
of this affirmation for what concerns those null modes with $l=0,1$: if one
mixes carefully $v_{1,-3,l=0,1}^{\left( 0\right) }$ and $v_{2,-3,l=0,1}^{%
\left( 0\right) }$ one can reach a suitably fine-tuned function that is in $%
\ker D$ and can be integrated in $d\tilde k$ with $\phi \left( k\right) $:
the fact that this fine tuned mixing cannot anyway be stable under Lorentz
group rules out this possibility, and it can be seriously stated that a
scalar product of $\phi \left( k\right) $ with the elements of $\ker D$ does
not exist.

Nevertheless there is a more subtle consequence of the existence of the null
modes of $D$ on the phase variable $\phi \left( k\right) $. The definition (%
\ref{5.111}) and Eq.(\ref{phi.5}) show that $\phi \left( k\right) $ has a
contribution from $\ker D$ consisting of the term $\left( k\cdot X\right) $,
that is from the solution $\left\{ v_{1,-3,l,m}^{\left( 0\right) }\right\} $
with $l=0,1$, but not from terms with $l\geq 2$, and this is clearly a
restriction on the mathematically possible configurations of $\phi \left(
k\right) $.

On the other hand, since the $\lambda =-3$ representation is indecomposable,
Minkowskian products like $\left( k\cdot X\right) $ are the only
Lorentz-invariant bilinear quantities which can be built up by involving a
finite number of elements of $\ker D$, because the $l=0,1$ subspace is the
only Lorentz-invariant subspace within $\ker D$. The construction of
Lorentz-invariant bilinear quantities depending on the whole set of
solutions could only be performed introducing series whose convergence
remain far from obvious, and are not under control.

\section{Conclusions and outlook.}

We have outlined an analysis of the problem of how to define collective and
relative canonical coordinates for a real Klein-Gordon classical field in
Minkowski space.

While the collective coordinates, with the meaning given to them in the
introduction, can be almost easily defined, the definition of the relative
ones presented a serious difficulty. To obtain a sound definition of these
variables we have found it necessary to put a set of restrictions on the
field, which are essentially equivalent to the vanishing of the so called
supertranslations.

We confined the discussion to the classical case. The problem of
quantization will face well known difficulties \cite{LYN}, and it will
require a separate work. In any case it is known that what can be well
defined, as a quantum operator, is the phase factor $\exp (i\phi )$, and not
the phase itself. So not really prohibitive difficulties should appear.

The very next step should be to apply the present analysis to the field
defined on a space-like surface, following the line mentioned in Section 1,
which goes back to Dirac \cite{DIR}. We may expect some simplification when
working on a family of space-like surfaces foliating Minkowski space, since
there the problem of covariance of the canonical variables changes. In
particular, on a hyperplane determined by the rest frame of the total
momentum, the four vectors split in a scalar and an irreducible spin 1 part 
\cite{LUSALBA}.

This collective-relative variable separating technique will soon be
exploited in that context: even with a very different kind of covariance, an
easy application of the present results will single out the right
gauge-fixing conditions for the Dirac-Bergman reduction of that problem \cite
{MATELUSA}.

Other interesting developments will be the application of this method to
different\ relativistic Bose fields (Maxwell as well as Yang-Mills) and to
non-relativistic fields (Schr\"odinger classical field, or fields of
fluid-dynamics). It could be interesting even to work out the
collective-relative splitting basis for solitonic configurations, while the
problem of constructing a similar framework for spinor fields will need
special cares \cite{Bigazzi}.

Another aspect that should deserve some interest is the possibility of
switching on a self-interaction among the relative variables. If this self
interaction is completely integrable we could get an integrable model for a
self interacting scalar field. This should be a highly non local
self-interaction, but it could in principle provide interesting integrable
models in Minkowski space.

Even if these are all promising further developments, it's clear that before
starting we should understand better, from a more physical point of view,
the consistency conditions (\ref{X.1}) and (\ref{X.2}).

When conditions formally identical to Eq.(\ref{X.1}) appear in the context
of asymptotic General Relativity they do in order to render well defined the
light-like-infinity Poincar\'e group, allowing to work out consistently the
ADM spin of some gravitational source (a pulsing star, a black hole and so
on).

Here the BMS charges $P_{lm}$'s appear in an unexpected way, making the
problem of understanding their presence even more complicated, and few
things about them seem to be clear: first, they must be constrained as in
Eq.(\ref{X.1}) in order for the transformations (\ref{quattro.3}) to be
consistent; second, they do not concern spacetime symmetries, and can simply
be considered generating functionals for canonical maps from an initial set
of Hamilton-Jacobi data into another \cite{bms-matelonghi}.

We have not been able to include these $P_{lm}$'s in a canonical framework,
we have not been able to find a basis in which they play the role of
canonical momenta conjugated to some generalized coordinates $X^{lm}$'s. We
have not been able to modify conditions (\ref{X.2}) in order to free the $%
P_{lm}$'s from constraints, because this should involve the definition of
suitable Lorentz-invariant bilinear quantities depending on the whole basis
of $\ker D$, and we have not been able to build up such quantities.

Anyway we have not an explicit no-go theorem obstructing the possibility of
working out a collective-relative splitting canonical set $\left\{ X^\mu
,P_\nu ;{\cal H},{\cal K};X^{lm},P_{l^{\prime }m^{\prime }}\right\} $, in
which the BMS charges are an essential part of the phase space.

At this stage we can only conjecture that, going on studying this particular
method of collective-relative splitting, either the no-go theorem will be
found (with an important meaning from a physical and group-theoretical point
of view), or the BMS-including phase space will be constructed.

\bigskip 

\acknowledgments 
The authors wish to thank L. Lusanna for many suggestions and discussions on
the subject of the present paper, for his useful criticism and for reading
the manuscript. The authors wish to thank G. Talenti and S. Marmi of the
Department of Mathematics of Florence too, for their useful advises and
suggestions.

\appendix


\section{Notations for the scalar field.}

We list here the various definitions concerning the scalar real Klein Gordon
field.

The Lagrangian and the Lagrangian density are

\begin{equation}
L=\int d^3x{\cal L},\quad \quad {\cal L}={\frac 12}\left( {\partial _\mu
\Phi \partial ^\mu \Phi }-m^2\Phi ^2\right) ;  \label{A.1}
\end{equation}

\noindent the field is real

\begin{equation}
{\overline{\Phi }}=\Phi ,  \label{A.2}
\end{equation}
(the bar means complex conjugate).

\noindent The canonical momentum and the equation of motion are

\begin{equation}
\Pi (x)={\dot \Phi }(x)\equiv {\partial _0\Phi (x)},\quad \quad (\Box
+m^2)\Phi (x)=0.  \label{A.3}
\end{equation}

The Noether current associated with a Lorentz transformation is

\begin{equation}
j^{\mu \nu }={\frac{\partial {\cal L}}{\partial \left( \partial _\mu \Phi
\right) }}\partial ^\nu \Phi -\eta ^{\mu \nu }{\cal L}=j^{\nu \mu },
\label{A.4}
\end{equation}

\begin{equation}
j^{\mu 0} = {\dot\Phi}\partial^{\mu}\Phi = \Pi\partial^{\mu}\Phi,\quad (\mu
\neq 0),  \label{A.5}
\end{equation}

\begin{equation}
j^{00}={\frac 12}\left[ {\dot \Phi }^2+\left( \vec \nabla \Phi \right)
^2+m^2\Phi ^2\right] .  \label{A.6}
\end{equation}

The Poincar\'e charges are

\begin{equation}
P^\mu =\int d^3xj^{0\mu }(x),\quad \quad M^{\mu \nu }=\int d^3x\left( x^\mu
j^{0\nu }-x^\nu j^{0\mu }\right) .  \label{A.7}
\end{equation}

For the existence of these generators it is sufficient that $\Phi (\vec x,\
.),\vec \nabla \Phi (\vec x,\ .)\in L_2({\bf R}^3)$.

The metric signature is $\eta\equiv (+,-,-,-)$.

The Poisson brackets are

\begin{equation}
\left\{ \Phi \left( \vec x,x^0\right) ,\Pi \left( \vec x^{\prime
},x^0\right) \right\} =\delta ^3\left( \vec x-\vec x^{\prime }\right) ,
\label{A.8}
\end{equation}

\noindent and the other equal-$x^0$ Poisson brackets vanish.

The Poincar\'e algebra is

\begin{equation}
\{P^{\mu},\ P^{\nu}\} = 0,  \label{A.9}
\end{equation}

\begin{equation}
\{M^{\mu \nu },\ P^\rho \}=P^\nu \eta ^{\mu \rho }-P^\mu \eta ^{\nu \rho },
\label{A.10}
\end{equation}

\begin{eqnarray}
\{M^{\mu \nu },\ M^{\rho \lambda }\} &=&-M^{\mu \lambda }\eta ^{\nu \rho
}-\eta ^{\mu \lambda }M^{\nu \rho }+  \nonumber \\
&&+M^{\mu \rho }\eta ^{\nu \lambda }+\eta ^{\mu \rho }M^{\nu \lambda }.
\label{A.11}
\end{eqnarray}

The Fourier transform of the field is

\begin{equation}
\Phi (\vec x,t)=\int \tilde {dk}[a(k)e^{-i(k\cdot x)}+c.c.],  \label{A.12}
\end{equation}

\noindent where c.c. means ''the complex conjugate'', and $(\ \cdot \ )$ is
the usual scalar product between 4-vectors. The component $k_0$ equals $%
\sqrt{\left| \vec k\right| ^2+m^2}$, and $\tilde {dk}$ is the Lorentz
invariant measure

\begin{eqnarray}
\tilde {dk} &=&{\frac{d^3k}{\Omega (k)}},\quad \Omega (k)=(2\pi )^32\omega
(k),  \nonumber \\
\quad \omega (k) &=&k_0=\sqrt{\left| \vec k\right| ^2+m^2}.  \label{A.13}
\end{eqnarray}

For the existence of the Fourier transform of the field $\Phi (x)$ we may
require that $a(k),\vec \nabla _ka(k)\in L_2(\tilde {dk})$.

The canonical momentum is

\begin{equation}
\Pi (x)=-i\int \tilde {dk}\omega (k)[a(k)e^{-i(k\cdot x)}-c.c.].
\label{A.14}
\end{equation}

In terms of the Fourier coefficients, the current $j^{00}$ is 
\begin{equation}
\begin{array}{l}
j^{0,0}(x)=%
\displaystyle \int 
\tilde {dk}%
\displaystyle \int 
\tilde {dk^{\prime }}\left\{ -\left[ \omega (k)\omega (k^{\prime })+\vec k%
\cdot \vec {k^{\prime }}-m^2\right] \left[ a(k)a(k^{\prime
})e^{-i(k+k^{\prime }\cdot x)}+c.c.\right] \right. + \\ 
\\ 
+\left. \left[ \omega (k)\omega (k^{\prime })+\vec k\cdot \vec {k^{\prime }}%
+m^2\right] \left[ a(k)\overline{a}(k^{\prime })e^{-i(k-k^{\prime }\cdot
x)}+c.c.\right] \right\} .
\end{array}
\label{A.141}
\end{equation}

The Poisson brackets are

\begin{equation}
\{a(k),\overline{a}(k^{\prime })\}=-i\Omega (k)\delta ^3(\vec k-\vec k%
^{\prime }).  \label{A.15}
\end{equation}

The inverse Fourier transform is

\begin{equation}
a(k)=\int d^3xe^{i(k\cdot x)}[\omega (k)\Phi (x)+i\Pi (x)].  \label{A.16}
\end{equation}

In terms of the Fourier coefficients, the Poincar\'e generators are

\begin{equation}
P^\mu =\int \tilde {dk}k^\mu \overline{a}(k)a(k),  \label{A.17}
\end{equation}

\begin{equation}
M_{ij}=M_{ij}^{\prime }=-i\int \tilde {dk}\overline{a}(k)\left( k^i{\frac 
\partial {\partial k^j}}-k^j{\frac \partial {\partial k^i}}\right) a(k),
\label{A.18}
\end{equation}

\begin{equation}
M_{0j}=x_0P_j+M_{0j}^{\prime }=x_0P_j+i\int \tilde {dk}\overline{a}(k)\omega
(k){\frac \partial {\partial k^j}}a(k),  \label{A.19}
\end{equation}

\noindent where $x_0=t$ is the parameter time, and $k^i=-k_i$.

\bigskip


\section{Some useful Poisson brackets.}

Let us verify the Eq.(\ref{1.39}) for $\mu =i$ and $\nu =j$: 
\[
\left\{ M_{ij}^{\prime },X_\nu \right\} =-\left( \eta _{i\nu }X_j-\eta
_{j\nu }X_i\right) . 
\]
From the Poisson brackets (\ref{1.3}) and the definitions (\ref{1.7}) and (%
\ref{1.20}) we get 
\[
\begin{array}{l}
\left\{ M_{ij}^{\prime },X_\nu \right\} =-%
\displaystyle \int 
d\tilde k%
{\displaystyle {\partial \over \partial P^\nu }}
F\left( P,k\right) \left( k_i%
{\displaystyle {\partial \over \partial k^j}}
-k_j%
{\displaystyle {\partial \over \partial k^i}}
\right) \phi \left( k\right) + \\ 
\\ 
+%
\displaystyle \int 
d\tilde k%
\displaystyle \int 
d\tilde k^{\prime }I\left( k\right) \phi \left( k^{\prime }\right) 
{\displaystyle {\partial ^2 \over \partial P^\rho \partial P^\nu }}
F\left( P,k^{\prime }\right) \left( k_i%
{\displaystyle {\partial \over \partial k^j}}
-k_j%
{\displaystyle {\partial \over \partial k^i}}
\right) k^\rho .
\end{array}
\]

Since we assumed a good behaviour of the function $F$, we may integrate by
parts the first integrand, and use Eq.(\ref{1.37}): 
\[
\begin{array}{l}
\left\{ M_{ij}^{\prime },X_\nu \right\} =-%
\displaystyle \int 
d\tilde k\phi \left( k\right) 
{\displaystyle {\partial \over \partial P^\nu }}
\left( P_i%
{\displaystyle {\partial \over \partial P^j}}
-P_j%
{\displaystyle {\partial \over \partial P^i}}
\right) F\left( P,k\right) + \\ 
\\ 
+\left( P_i%
{\displaystyle {\partial \over \partial P^j}}
-P_j%
{\displaystyle {\partial \over \partial P^i}}
\right) 
\displaystyle \int 
d\tilde k\phi \left( k\right) 
{\displaystyle {\partial \over \partial P^\nu }}
F\left( P,k\right) =-\left( \eta _{i\nu }X_j-\eta _{j\nu }X_i\right) ,
\end{array}
\]
where there is a compensation of the last term.

In the same way, using Eq.(\ref{1.38}) and taking care of the mass shell
condition $k^0=\omega \left( k\right) $ in the calculation of $\omega \left(
k\right) \frac \partial {\partial k^j}k^\rho $, we get 
\[
\left\{ M_{0j}^{\prime },X_\nu \right\} =-\left( \eta _{0\nu }X_j-\eta
_{j\nu }X_0\right) . 
\]

\bigskip


\section{The Laplace-Beltrami operator.}

The Laplace-Beltrami operator of Eq.(\ref{2.20}) has been studied in a
series of papers by Raczka, Limic and Niederle \cite{RLN}, where it is
called $-\Delta (H_3^1)$ (see Eq.(5.10) in the first reference). There it is
shown that $\Delta $ has not a discrete spectrum, but only a continuous one,
and the basis of its generalized eigenfunctions is determined. That study is
the background of our work \cite{bms-matelonghi} too, where the reader will
find a complete discussion of the spectrum and of the properties of this
operator.

The zero modes of the operator (\ref{2.25}) correspond to the

\begin{equation}
\lambda =-3  \label{3.14}
\end{equation}
representation of the notations used in \cite{bms-matelonghi}, that is a non
unitary representation of the Lorentz group, which is reducible but not
completely reducible as we have shown in \cite{bms-matelonghi}.

\noindent The value $\lambda =-3$ is determined by the conditions, Eq.(\ref
{2.1})

\begin{equation}
\int \tilde {dk}k^\mu \hat I(k)=0,\quad \hat I(k)=D{\cal H}(k),
\label{3.15bis}
\end{equation}

\noindent which was assumed as defining the relative variable ${\cal H}$.
This choice is reinforced by the fact that this is the unique eigenspace of $%
\Delta $ which allows an invariant subspace with dimension 4 (see \cite
{bms-matelonghi}), corresponding to a 4-vector, that is $k^\mu $: so the
requirement of the existence of a 4-vector $P^\mu $ as in Eq.(\ref{1.1})
among our variables selects by itself the choice $\lambda =-3$.

Working on the functional space of 
\[
f:H_3^1\longmapsto {\bf C} 
\]
on which $\Delta $ and $D$ act, the generators of $SO\left( 1,3\right) $ are
represented by the following differential operators 
\begin{equation}
\begin{array}{l}
l_{\mu \nu }=iD_{\mu \nu }, \\ 
l_{ij}=-\epsilon _{ijk}L_k=i\left( x^i\partial _j-x^j\partial _i\right) , \\ 
l_{0j}=-K_j=-i\sqrt{1+\left| \vec x\right| ^2}\ \partial _j,
\end{array}
\label{3.15}
\end{equation}
where

\begin{equation}
\begin{array}{cc}
\vec x={%
{\displaystyle {\vec k \over m}}
,} & r=\left| \vec x\right|
\end{array}
\label{3.16bis}
\end{equation}
and $D_{\mu \nu }$ is the operator we defined in Eq.(\ref{1.32}).

$D$ reads

\begin{equation}
D=-m^2\Delta +3=-\Delta |_x+3,  \label{3.17}
\end{equation}

\noindent where

\begin{equation}
\Delta |_x=\left( 1+r^2\right) {\frac{\partial ^2}{\partial r^2}}+\left( {%
\frac 2r}+3r\right) {\frac \partial {\partial r}}-{\frac{J^2}{r^2}}
\label{3.18}
\end{equation}

\noindent and $J^2=\left| \vec L\right| ^2$ as usually.

Its proper eigenfunctions correspond to the values of $\lambda \ \in \
[1,+\infty )$, or $\Lambda \ \in \ [0,+\infty )$ where 
\begin{equation}
\Lambda =\sqrt{\lambda -1},  \label{b.3.18}
\end{equation}
see Appendix D, taken as a limit from the upper complex half-plane of $%
\Lambda $ see \cite{RLN}. They read:

\begin{equation}
\begin{array}{rcl}
w_{\lambda ,l,m}(r,\theta ,\phi ) & = & N_{\lambda l}\ v_{1,\lambda
,l,m}^{(0)}= \\ 
&  &  \\ 
& = & N_{\lambda l}\ r^l\ _2F_1\left( {%
{\displaystyle {l+i+i\Lambda  \over 2}}
},{%
{\displaystyle {l+i-i\Lambda  \over 2}}
};l+{%
{\displaystyle {3 \over 2}}
};-r^2\right) Y_{l,m}(\theta ,\phi ).
\end{array}
\label{3.19}
\end{equation}
Here $_2F_1$ is the hypergeometric function, $l=0,1,2,...$, and $\left|
m\right| \leq l$ and the normalization factor $N_{\lambda l}$ is

\begin{equation}
N_{\lambda l} = {\frac{2\pi}{m\sqrt{\Lambda}}}\left\vert{\frac{\Gamma({\frac{%
l+2-i\Lambda}{2}}) \Gamma({\frac{l+1-i\Lambda}{2}})}{{\Gamma(i\Lambda)%
\Gamma(l+{\frac{3}{2}})}}} \right\vert.  \label{3.21}
\end{equation}

With this normalization $w_{\lambda ,l,m}$ is an orthonormal set with
respect to the scalar product of $L_2(\tilde {dk})$

\begin{equation}
\int \tilde {dk}\bar w_{\lambda lm}(r,\theta ,\phi )w_{\lambda
,l,m}(r^{\prime },\theta ^{\prime },\phi ^{\prime })=\delta _{ll^{\prime
}}\delta _{mm^{\prime }}\delta (\Lambda -\Lambda ^{\prime }).  \label{3.22}
\end{equation}

This basis is complete in $L_2(\tilde{dk})$, that is

\begin{equation}
\sum_{l\geq 0}\sum_{\mid m\mid \leq l}\int_1^\infty d\lambda w_{\lambda
,l,m}(r,\theta ,\phi )\bar w_{\lambda ,l,m}(r^{\prime },\theta ^{\prime
},\phi ^{\prime })=\Omega (k)\delta ^3(k-k^{\prime }).  \label{3.30}
\end{equation}

From now on we define

\begin{equation}
v_{\lambda ,l,m}(r,\theta ,\phi )=v_{1,\lambda ,l,m}^{(0)}=r^l\ _2F_1\left( {%
\frac{l+i+i\Lambda }2},{\frac{l+i-i\Lambda }2};l+{\frac 32};-r^2\right)
Y_{l,m}(\theta ,\phi ).  \label{3.31}
\end{equation}
The expressions written up to now hold for $\Lambda \ \in \ [0,+\infty )$,
or $\lambda \ \in \ [1,+\infty )$, which corresponds to the continuous
spectrum of $\Delta $. Since we are interested the case $\lambda =-3$, we
need the same relations for a generic complex $\lambda $.

For $\lambda =-3$ the normalization factor $N_{\lambda l}$ becomes zero for $%
l\geq 2$, moreover the functions $w_{\lambda ,l,m}$ are no more normalizable.

\subsection{The space $\ker \left( \Delta +\lambda \right) $.}

Let us study the solutions of the equation

\begin{equation}
\left( \Delta +\lambda \right) f=0,  \label{B.1}
\end{equation}
for generic values of $\lambda \in {\bf C}$.

A fundamental system of solutions, in the neighbourhood of the origin, that
is for $r\simeq 0$, of the Eq.(\ref{B.1}), is

\begin{eqnarray}
v_{1,\lambda ,l,m}^{(0)}(\vec r) &=&u_{1,\lambda ,l}^{(0)}(r)Y_{l,m}(\theta
,\phi ),  \nonumber \\
v_{2,\lambda ,l,m}^{(0)}(\vec r) &=&u_{2,\lambda ,l}^{(0)}(r)Y_{l,m}(\theta
,\phi ),  \label{B.2}
\end{eqnarray}
that is: 
\begin{equation}
\left\{ 
\begin{array}{l}
u_{1,\lambda ,l}^{(0)}(r)=r^l\ _2F_1\left( {%
{\displaystyle {l+1+i\Lambda  \over 2}}
},{%
{\displaystyle {l+1-i\Lambda  \over 2}}
};l+{%
{\displaystyle {3 \over 2}}
};-r^2\right) , \\ 
\\ 
u_{2,\lambda ,l}^{(0)}(r)=r^{-l-1}\ _2F_1\left( -{%
{\displaystyle {l+i\Lambda  \over 2}}
},-{%
{\displaystyle {l-i\Lambda  \over 2}}
};{%
{\displaystyle {1 \over 2}}
}-l;-r^2\right) , \\ 
\\ 
u_{2,\lambda ,l}^{(0)}(r)=u_{1,\lambda ,-l-1}^{(0)}(r).
\end{array}
\right.  \label{B.3}
\end{equation}

A fundamental system in the neighborhood of the point at infinite $%
r\rightarrow \infty $ is

\begin{eqnarray}
v^{(\infty)}_{1,\lambda,l,m}(\vec{r}) = u_{1,\lambda,l}^{(\infty)}(r)
Y_{l,m}(\theta,\phi),  \nonumber \\
v^{(\infty)}_{2,\lambda,l,m}(\vec{r}) = u_{2,\lambda,l}^{(\infty)}(r)
Y_{l,m}(\theta,\phi),  \label{B.4}
\end{eqnarray}

\noindent where 
\begin{equation}
\left\{ 
\begin{array}{c}
u_{1,\lambda ,l}^{(\infty )}(r)=r^{-1-i\Lambda }\ _2F_1\left( 
{\displaystyle {l+1+i\Lambda  \over 2}}
,-%
{\displaystyle {l-i\Lambda  \over 2}}
;l+i\Lambda ;-{%
{\displaystyle {1 \over r^2}}
}\right) , \\ 
\\ 
u_{2,\lambda ,l}^{(\infty )}(r)=r^{-1+i\Lambda }\ _2F_1\left( 
{\displaystyle {l+1-i\Lambda  \over 2}}
,-%
{\displaystyle {l+i\Lambda  \over 2}}
;1-i\Lambda ;-{%
{\displaystyle {1 \over r^2}}
}\right) .
\end{array}
\right.  \label{B.5}
\end{equation}

For $r\geq 0$ no other singular points are met. For $z=-1$ the
hypergeometric series $_2F_1\left( {\alpha },{\beta };\gamma ;z\right) $ is
absolutely convergent, since its coefficients satisfy

\begin{equation}
\mathop{\rm Re}
(\alpha +\beta -\gamma )=-{\frac 12}.  \label{B.6}
\end{equation}

For $i\Lambda $ integer positive, the solution $u_{1,\lambda ,l}^{(\infty
)}(r)$ should be modified, but since we will be interested in the solution $%
u_{2,\lambda ,l}^{(\infty )}(r)$ for ${\rm Im}\Lambda \geq 0$, we will not
give here the necessary modification.

We are interested in the normalization properties of these solutions in the
neighborhood of the point $0$ and $\infty$, with respect to the invariant
measure

\begin{equation}
\tilde{dr} = {\frac{d^3 r}{\sqrt{1+r^2}}}.  \label{B.7}
\end{equation}

The solution $u_{1,\lambda ,l,m}^{(0)}$ is {\it regular and normalizable in
the neighborhood of the origin}. The solution $u_{2,\lambda ,l,m}^{(0)}$ on
the other hand is {\it normalizable in the origin for }$l=0${\it \ only}. We
will discard this solution, even in the case $l=0$, since under the action
of a boost it would be transformed in a solution with a different value of $%
l $, that is non normalizable \cite{bms-matelonghi}.

The solution $u_{1,\lambda ,l}^{(0)}$ can be analytically continued to $%
r\rightarrow \infty $%
\begin{equation}
\begin{array}{l}
u_{1,\lambda ,l}^{(0)}(r)= \\ 
\\ 
= 
{\displaystyle {\Gamma \left( \frac 32+l\right) \Gamma \left( -i\Lambda \right)  \over \Gamma \left( {\frac{l+1-i\Lambda }2}\right) \Gamma \left( \frac l2+1-\frac{i\Lambda }2\right) }}
r^{-1-i\Lambda }\ _2F_1\left( {%
{\displaystyle {l+1+i\Lambda  \over 2}}
},{%
{\displaystyle {i\Lambda -l \over 2}}
};1+i\Lambda ;-r^{-2}\right) + \\ 
\\ 
+ 
{\displaystyle {\Gamma \left( \frac 32+l\right) \Gamma \left( i\Lambda \right)  \over \Gamma \left( {\frac{l+1+i\Lambda }2}\right) \Gamma \left( \frac l2+1+\frac{i\Lambda }2\right) }}
r^{-1+i\Lambda }\ _2F_1\left( {%
{\displaystyle {l+1-i\Lambda  \over 2}}
},-{%
{\displaystyle {i\Lambda +l \over 2}}
};1-i\Lambda ;-r^{-2}\right)
\end{array}
\label{B.7.bis}
\end{equation}
and, for a generic value of $\Lambda $ has a behaviour which is a linear
combination of the two power of $r$, see Eq. 2.10(2) of \cite{BAT}

\begin{equation}
r^{-1-i\Lambda },\quad r^{-1+i\Lambda },  \label{B.8}
\end{equation}

\noindent which, for a real $\Lambda $, is normalizable.

For a real $\Lambda $ this is an eigenfunction of the operator $\Delta $,
see \cite{RLN}, belonging to the continuous spectrum. Since $\lambda
=1+\Lambda ^2$, the spectrum corresponds to $\lambda \ \in \ [1,+\infty )$.
The upper half-plane of the complex $\Lambda $ plane describes all the
values of $\lambda $, and the spectrum is obtained as a limit $\lambda =y+i0$%
, with $y\ \in \ [1,+\infty )$.

For the second fundamental system of Eq.(\ref{B.5}), we have that the
solution $u_{1,\lambda ,l}^{(\infty )}$ has the behaviour, for $r\simeq
\infty $

\begin{equation}
r^{-1+{\rm Im}\Lambda-i{\rm Re}\Lambda},  \label{B.10}
\end{equation}

\noindent and, for ${\rm Im}\Lambda \geq 0$, is {\it not normalizable}.

The solution $u_{2,\lambda ,l,m}^{(\infty )}$ is instead {\it normalizable}
for $r\rightarrow \infty $ and it will be used in the next Appendix,
together with the solution $u_{1,\lambda ,l,m}^{(0)}$, for the determination
of the Green function of the $D$ operator. The behaviour of the solutions (%
\ref{B.4}) is exhibited in reference \cite{bms-matelonghi}.

\subsection{The $\lambda =-3$ solutions.}

In the case $\lambda =-3$, that is $\Lambda =2i$, we must use another
asymptotic expansion. More in general, for $\Lambda =ni$, with $n$ integer,
we must use the expansion given in Eq. 2.10(7) of \cite{BAT}; we get, for $%
\Lambda =2i$ and for $r\rightarrow \infty $

\begin{equation}
u_{1,-3,l}^{(0)}\simeq r+O(r^{-1}).  \label{B.9}
\end{equation}
So, this solution is no more normalizable at $\infty $, but only in $0$.

The other solution has a behaviour

\begin{equation}
u_{2,-3,l}^{(0)}\simeq r+O(r^{-1}),\quad {\rm for}\quad l=0,1;\quad
u_{2,-3,l}^{(0)}\simeq B_lr^{-3},\quad {\rm for}\quad l\geq 2.\quad
\label{B.91}
\end{equation}
and is singular in the origin as $r^{-l-1}$ (in (\ref{B.91}) the coefficient 
$B_l$ is the same as in (\ref{C.51}), see below).

Let us summarize the fundamental solutions around $r=0$ and $r\rightarrow
\infty $ in the particular case of $\Lambda =2i$, the case we are interested
in.

The fundamental system of solutions around $r=0$ is 
\begin{equation}
v_{1,-3,l}^{(0)}(\vec r)=r^l\ _2F_1\left( \frac{l-1}2,\frac{l+3}2;l+{\frac 32%
};-r^2\right) Y_{l,m}(\theta ,\phi ),  \label{B.91.a}
\end{equation}

\begin{eqnarray}
v_{2,-3,l}^{(0)}(\vec r)=r^{-l-1}\ _2F_1\left( \frac{2-l}2,-\frac{l+2}2;{%
\frac 12}-l;-r^2\right) Y_{l,m}(\theta ,\phi ).  \label{B.91.b}
\end{eqnarray}

The fundamental system of solutions around $r\rightarrow \infty $ for $%
\lambda =-3$ reads: 
\begin{equation}
v_{1,-3,l}^{(\infty )}(\vec r)=r\ _2F_1\left( \frac{l-1}2,-\frac{l+2}2;l-2;-{%
\frac 1{r^2}}\right) Y_{l,m}(\theta ,\phi ),  \label{B.91.c}
\end{equation}
\begin{equation}
v_{2,\lambda ,l}^{(\infty )}(\vec r)=r^{-3}\ _2F_1\left( \frac{l+3}2,\frac{%
2-l}2;3;-{\frac 1{r^2}}\right) Y_{l,m}(\theta ,\phi ).  \label{B.91.d}
\end{equation}

Asymptotic behaviours for these $v$'s are those shown in the previous
subsection, adapted as $\Lambda =2i$.

\bigskip


\section{Green function of the operator $D$.}

In this Appendix we study the Green function $G(\vec k,\vec k^{\prime
};\lambda )$ of the operator $\Delta $ for a generic value of $\lambda $,
belonging to the complex plane cut along the real axis in $\lambda \ \in \
[1,+\infty )$. As in Appendix C it is useful to define $\lambda =1+\Lambda
^2 $, with ${\rm Im}\Lambda \geq 0$. Then we put $\lambda =-3$ finding out
the particular case we are interested in.

\subsection{Generic complex $\lambda $.}

We define the distribution $G$ as the solution of the equation

\begin{equation}
\left( -m^2\Delta _k-\lambda \right) G(\vec k,\vec k^{\prime };\lambda
)=\Omega (k)\delta ^3(\vec k-\vec k^{\prime }),  \label{C.1}
\end{equation}

\noindent where $\Delta$ is defined in Eq.(\ref{2.20}) or Eq.(\ref{3.18}),
in terms of the variable $\vec{x} = {\frac{\vec{k}}{m}}$.

The solutions of the homogeneous equation have been studied in Appendix C.
Following the usual procedure \cite{Stakgold}, since the origin and the
point at $\infty $ are regular singular points for the radial equation, we
will choose the solutions $v_{1,\lambda ,l,m}^{(0)}$ and $v_{2,\lambda
,l,m}^{(\infty )}$, which are normalizable in the origin and in the point at
infinity respectively. So we will write the Green function as

\begin{equation}
G(\vec{k},\vec{k}^{\prime};\lambda) = \sum_{l\geq 0}\sum_{\mid m\mid\leq l} {%
Y}_{l,m}(\hat{k}) \overline{Y}_{l,m}(\hat{k}^{\prime}) {\cal {G}}%
_l(r,r^{\prime};\lambda),  \label{C.2}
\end{equation}

\noindent where $\hat k$ and $\hat k^{\prime }$ are the directions of $\vec k
$ and $\vec k^{\prime }$, where

\begin{equation}
r = {\frac{1}{m}} \mid\vec{k}\mid,\quad r^{\prime} = {\frac{1}{m}} \mid\vec{k%
}^{\prime}\mid,  \label{C.3}
\end{equation}

\noindent and where ${\cal {G}}$ is the radial Green function given by

\begin{equation}
{\cal {G}}_l(r,r^{\prime };\lambda )={\cal A}_l(\Lambda )\left[ u_{1,\lambda
,l}^{(0)}(r)u_{2,\lambda ,l}^{(\infty )}(r^{\prime })\theta (r^{\prime
}-r)+(r\leftrightarrow r^{\prime })\right] .  \label{C.4}
\end{equation}
The Green function $G(\vec k,\vec k^{\prime };\lambda )$ is symmetric as it
should be since the operator $\Delta $ is self-adjoint.

The radial functions $u_{1,\lambda ,l}^{(0)}(r)$ and $u_{2,\lambda
,l}^{(\infty )}(r^{\prime })$ are defined in Eq.(\ref{B.3}) and Eq.(\ref{B.5}%
): 
\begin{equation}
u_{1,\lambda ,l}^{(0)}(r)=r^l\ _2F_1\left( {\frac{l+1+i\Lambda }2},{\frac{%
l+1-i\Lambda }2};l+{\frac 32};-r^2\right) ,  \label{C.5.prima}
\end{equation}
\begin{equation}
u_{2,\lambda ,l}^{(\infty )}(r^{\prime })=r^{-3}\ _2F_1\left( {\frac{%
l+1-i\Lambda }2},-{\frac{l+i\Lambda }2};1-i\Lambda ;-{\frac 1{r^2}}\right) ,
\label{C.5}
\end{equation}

\noindent and their asymptotic behaviour was discussed in Appendix C.

The action of the Lorentz generators on the solution $v_{1,\lambda
,l,m}^{(0)}$ has been studied in \cite{bms-matelonghi}. The action on the
solution $v_{2,\lambda ,l,m}^{(\infty )}$ can be obtained from the relation

\begin{equation}
u_{2,\lambda ,l}^{(\infty )}=A_l\ u_{1,\lambda ,l}^{(0)}(r)+B_l\
u_{2,\lambda ,l}^{(0)}(r),  \label{C.51}
\end{equation}

\noindent where

\begin{equation}
A_l={\frac{\Gamma (1-i\Lambda )\Gamma \left( -l-{\frac 12}\right) }{\Gamma
\left( -{\frac{i\Lambda +l}2}\right) \Gamma \left( {\frac{1-i\Lambda -l}2}%
\right) }},  \label{C.52}
\end{equation}

\begin{equation}
B_l={\frac{\Gamma (1-i\Lambda )\Gamma \left( l+{\frac 12}\right) }{\Gamma
\left( {\frac{l+1-i\Lambda }2}\right) \Gamma \left( {\frac{2-i\Lambda +l}2}%
\right) }},  \label{C.53}
\end{equation}

\noindent with, for $\Lambda =2i$, $A_l=0$ if $l\geq 2$.

The radial Green function ${\cal {G}}$ satisfies the equation

\begin{equation}
(-\Delta _l-\lambda ){\cal {G}}_l(r,r^{\prime };\lambda )=\Omega (k){\frac{%
\delta (r-r^{\prime })}{m^3r^2}},  \label{C.6}
\end{equation}

\noindent where the operator $\Delta_l$ is given in Eq.(\ref{3.18}), with $%
J^2\rightarrow l(l+1)$, and where

\begin{equation}
\Omega(k) = 2(2\pi)^3 m \sqrt{1+r^2}.  \label{C.7}
\end{equation}

Since ${\rm Im}\Lambda\geq 0$, the characteristic exponent at the point at $%
\infty$ of the solution $u_{2,\lambda,l}^{\infty}(r)$ is (in ${\frac{1}{r}}$%
) given by

\begin{equation}
1-i\Lambda = 1+Im\Lambda - iRe\Lambda,  \label{C.8}
\end{equation}

\noindent and since the other solution $u_{1,\lambda,l}^{\infty}$ has (in ${%
\frac{1}{r}}$) the exponent

\begin{equation}
1+i\Lambda = 1 - Im\Lambda + iRe\Lambda ,  \label{C.9}
\end{equation}

\noindent see the Appendix C, Eq.(\ref{B.10}), it follows that the
logarithmic case does not appear.

The constant ${\cal A}_l(\Lambda )$ in the Eq.(\ref{C.4}) is given by

\begin{equation}
{\cal A}_l(\Lambda )=-{\frac 1{(1+r^2)W_l(r)}}{\frac{\Omega (k)}{m^3r^2}},
\label{C.10}
\end{equation}

\noindent where $W_l(r)$ is the Wronskian

\begin{equation}
W_l(r)=u_{1,\lambda ,l}^{(0)}(r){\frac d{dr}}u_{2,\lambda ,l}^{(\infty
)}(r)-u_{2,\lambda ,l}^{(\infty )}(r){\frac d{dr}}u_{1,\lambda ,l}^{(0)}(r),
\label{C.11}
\end{equation}

\noindent which turns out to be

\begin{equation}
W_l(r) = {\frac{K}{r^2\sqrt{1+r^2}}},  \label{C.12}
\end{equation}

\noindent with

\begin{equation}
K = -2{\frac{\Gamma(l+{\frac{3}{2}})\Gamma(1-i\Lambda)}{\Gamma({\frac{%
l+2-i\Lambda}{2}})\Gamma({\frac{l+1-i\Lambda}{2}})}}.  \label{C.13}
\end{equation}

Collecting everything we have

\begin{equation}
{\cal A}_l(\Lambda )=-{\frac{(2\pi )^3}{m^2}}{\frac{\Gamma ({\frac{%
l+2-i\Lambda }2})\Gamma ({\frac{l+1-i\Lambda }2})}{\Gamma (l+{\frac 32}%
)\Gamma (1-i\Lambda )}}.  \label{C.14}
\end{equation}
The equations (\ref{C.2}), (\ref{C.4}), (\ref{C.5.prima}), (\ref{C.5}) and (%
\ref{C.14}) give the complete expression of the Green function $G(\vec k,%
\vec k^{\prime };\lambda )$.

The so expressed Green function is an analytic function of $\lambda $ in the
complex plane cut along the real axis for $\lambda \ \in \ [0,+\infty )$; it
vanishes for $\mid \lambda \mid \rightarrow \infty $, when $r\not =r^{\prime
}$. Its asymptotic behaviour is given by

\begin{equation}
{\cal {G}}_l(r,r^{\prime};\lambda) = -{\frac{(2\pi)^3}{m^2}}{\frac{1}{r
r^{\prime}}} {\frac{1}{\Lambda}}\left[ e^{{\frac{i}{2}}\mid\zeta^{\prime}-%
\zeta\mid\Lambda} + (-1)^{l+1} e^{{\frac{i}{2}}(\zeta^{\prime}+\zeta)%
\Lambda}\right],  \label{C.15}
\end{equation}

\noindent where

\begin{equation}
\zeta = 2\ln(r+\sqrt{1+r^2}),\quad \zeta^{\prime} = 2\ln(r^{\prime}+\sqrt{%
1+r^{\prime 2}}).  \label{C.16}
\end{equation}

In the Eq.(\ref{C.15}) use was made of the Eq.2.3.2(16),(17) of the
reference \cite{BAT}.

From the Eq.(\ref{C.15}) we see that ${\cal {G}}$ is exponentially vanishing
when $\mid\lambda\mid\rightarrow\infty$ for $r \not= r^{\prime}$. In the
limit $r^{\prime}\rightarrow r$ we have

\begin{equation}
{\frac{1}{2\pi i}}\int_{C_{\infty}} d\lambda {\cal {G}}_l(r,r^{\prime};%
\lambda) = 2(2\pi)^3{\frac{\sqrt{1+r^2}}{m^2 r^2}}\delta(r-r^{\prime}),
\label{C.17}
\end{equation}

\noindent where $C_{\infty}$ is a path of integration at infinity, which
avoids the positive real axis. The last equation is useful for the
demonstration of the completeness of the eigenfunctions of the
Laplace-Beltrami operator.

We have indeed, from the Cauchy theorem

\begin{equation}
{\frac 1{2\pi \imath }}\int_{C_\infty }{\cal {G}}(r,r^{\prime };\lambda
)d\lambda =-{\frac 1{2\pi i}}\int_{{\rm cut}}{\cal {G}}(r,r^{\prime
};\lambda )d\lambda =-{\frac 1{2\pi i}}\int_1^\infty \left[ {\cal {G}}%
_l\right] \left( r,r^{\prime };\lambda \right) d\lambda ,  \label{C.18}
\end{equation}

\noindent where $\left[ {\cal {G}}_l\right]$ is the discontinuity across the
cut of ${\cal {G}}$. Explicitly

\begin{equation}
\begin{array}{c}
\left[ {\cal {G}}_l\right] (r,r^{\prime };\lambda )={\cal {G}}_l(r,r^{\prime
};\lambda +i0)-{\cal {G}}_l(r,r^{\prime };\lambda -i0)= \\ 
=-{\frac{8i\pi ^3}{m^2}}{\frac 1\Lambda }\left| {\frac{\Gamma ({\frac{%
l+2-i\Lambda }2})\Gamma ({\frac{l+11-i\Lambda }2})}{\Gamma (i\Lambda )\Gamma
(l+{\frac 32})}}\right| ^2 \\ 
r^l\ _2F_1({\frac{l+1+i\Lambda }2},{\frac{l+1-i\Lambda }2};l+{\frac 32}%
;-r^2)r^{\prime l}\ _2F_1({\frac{l+1+i\Lambda }2},{\frac{l+1-i\Lambda }2};l+{%
\frac 32};-r^{\prime 2}),
\end{array}
\label{C.19}
\end{equation}

\noindent where $\Lambda\ \in\ [0,+\infty)$.

If we define

\begin{equation}
w_{\lambda ,l,m}(\vec k)=N_{\lambda ,l}v_{1,\lambda ,l,m}^{(0)}(\vec r%
)=N_{\lambda ,l}r^l\ _2F_1\left( {\frac{l+1+i\Lambda }2},{\frac{l+1-i\Lambda 
}2};l+{\frac 32};-r^2\right) Y_{l,m}(\hat k),  \label{C.20}
\end{equation}

\noindent where $\hat{k}$ is the direction of $\vec{k}$, and

\begin{equation}
N_{\lambda,l} = {\frac{2\pi}{m\sqrt{\Lambda}}} \left\vert{\frac{\Gamma({%
\frac{l+2-i\Lambda}{2}})\Gamma({\frac{l+11-i\Lambda}{2}})}{%
\Gamma(i\Lambda)\Gamma(l+{\frac{3}{2}})}}\right\vert,  \label{C.21}
\end{equation}

\noindent we have the completeness of the eigenfunctions $w_{\lambda ,l,m}$

\begin{equation}
\sum_{l\geq 0}\sum_{\mid m\mid \leq 0}\int_1^\infty d\lambda w_{\lambda
,l,m}(\vec k)\bar w_{\lambda ,l,m}(\vec k^{\prime })=\Omega (k)\delta ^3(%
\vec k-\vec k^{\prime }).  \label{C.22}
\end{equation}

So we have recovered the result of \cite{RLN} by the use of the Green
function.

The eigenfunctions $w_{\lambda ,l,m}$ are an orthogonal set with respect to
the invariant measure $\tilde {dk}$

\begin{equation}
\int \tilde {dk}\bar w_{\lambda ,l,m}(\vec k)w_{\lambda ^{\prime },l^{\prime
},m^{\prime }}(\vec k)=\delta (\lambda -\lambda ^{\prime })\delta
_{l,l^{\prime }}\delta _{m,m^{\prime }}.  \label{C.23}
\end{equation}

The action of the Lorentz generators on this basis is studied in \cite
{bms-matelonghi}.

In terms of these eigenfunction the Green function has the expression

\begin{equation}
G(\vec k,\vec k^{\prime };\lambda )=\int_1^\infty d\lambda ^{\prime
}\sum_{lm}\frac{w_{\lambda ^{\prime },l,m}(\vec k){\bar w}_{\lambda ^{\prime
},l,m}(\vec {k^{\prime }})}{\lambda ^{\prime }-\lambda }.  \label{C.24}
\end{equation}

\subsection{The $\lambda =-3$ case.}

The Green function of the operator $D$ is obtained by putting $\lambda =-3$
in Eq.(\ref{C.4}), as

\begin{equation}
G(k, k^{\prime}) \equiv G(\vec{k},\vec{k}^{\prime};-3),  \label{C.25}
\end{equation}

\noindent which is real and symmetric. Here we give again the explicit
expressions of the radial parts of null modes of $\Delta -3$ whose
behaviours make them good to build up Green's functions as in (\ref{C.4}): 
\begin{equation}
\begin{array}{l}
u_{1,-3,l}^{(0)}(r)=r^l\ _2F_1\left( 
{\displaystyle {l-1 \over 2}}
,%
{\displaystyle {l+3 \over 2}}
;l+%
{\displaystyle {3 \over 2}}
;-r^2\right) , \\ 
\\ 
u_{2,-3,l}^{(\infty )}(r)=r^{-3}\ _2F_1\left( 
{\displaystyle {l+3 \over 2}}
,-%
{\displaystyle {l-2 \over 2}}
;3;-%
{\displaystyle {1 \over r^2}}
\right) .
\end{array}
\label{C.26}
\end{equation}
These are put together as follows: 
\begin{equation}
\begin{array}{l}
G\left( \vec k,\vec k^{\prime }\right) =%
\mathop{\displaystyle \sum }
\limits_{l\geq 0}%
\mathop{\displaystyle \sum }
\limits_{\mid m\mid \leq l}{Y}_{l,m}(\hat k)\overline{Y}_{l,m}(\hat k%
^{\prime }){\cal {G}}_l(r,r^{\prime };-3), \\ 
\\ 
{\cal {G}}_l(r,r^{\prime };-3)= \\ 
\\ 
=-{%
{\displaystyle {(2\pi )^3 \over m^2}}
} 
{\displaystyle {\Gamma \left( \frac{l+4}2\right) \Gamma \left( \frac{l+3}2\right)  \over \Gamma \left( l+\frac 32\right) \Gamma (3)}}
\left[ u_{1,-3,l}^{(0)}(r)u_{2,-3,l}^{(\infty )}(r^{\prime })\theta
(r^{\prime }-r)+u_{1,-3,l}^{(0)}(r^{\prime })u_{2,-3,l}^{(\infty )}(r)\theta
(r-r^{\prime })\right]
\end{array}
\label{C.27}
\end{equation}

\bigskip


\section{Lorentz generators and their algebra.}

In this Appendix we give some detail of the calculation of the algebra of
the Lorentz generators.

Let us limit ourselves to $M_{ij}$, which is given in Eq.(\ref{1.20})

\begin{equation}
M_{ij}=\int \tilde {dk}I(k)\left( k^i{\frac \partial {\partial k^j}}-k^j{%
\frac \partial {\partial k^i}}\right) \phi (k).  \label{D.1}
\end{equation}

Expressing $I(k)$ and $\phi (k)$ in terms of the auxiliary variables as in
Eq.(\ref{2.1}) and Eq.(\ref{2.2}) we get

\begin{equation}
\begin{array}{l}
M_{ij}=%
\displaystyle \int 
\tilde {dk}\hat I(k)\left( k^i{%
{\displaystyle {\partial \over \partial k^j}}
}-k^j{%
{\displaystyle {\partial \over \partial k^i}}
}\right) \hat \phi (k)+ \\ 
\\ 
+%
\displaystyle \int 
\tilde {dk}I(k)\left( k^i{%
{\displaystyle {\partial \over \partial k^j}}
}-k^j{%
{\displaystyle {\partial \over \partial k^i}}
}\right) (k\cdot X)+ \\ 
\\ 
+%
\displaystyle \int 
\tilde {dk}F((P\cdot k),P^2)\left( k^i{%
{\displaystyle {\partial \over \partial k^j}}
}-k^j{%
{\displaystyle {\partial \over \partial k^i}}
}\right) \hat \phi (k).
\end{array}
\label{D.2}
\end{equation}

The last term is zero: indeed, if we integrate by parts and remember the
asymptotic behaviour of $F$ for $\mid \vec k\mid \rightarrow \infty $, which
allows to neglect the surface term at infinity, we get, using Eq.(\ref{1.37})

\begin{equation}
\begin{array}{l}
-%
\displaystyle \int 
\tilde {dk}\hat \phi (k)\left( k^i{%
{\displaystyle {\partial \over \partial k^j}}
}-k^j{%
{\displaystyle {\partial \over \partial k^i}}
}\right) F((P\cdot k),P^2)= \\ 
\\ 
=-\left[ P^j%
\displaystyle \int 
\tilde {dk}\hat \phi (k){%
{\displaystyle {\partial F((P\cdot k),P^2) \over \partial P^i}}
}-(i\leftrightarrow j)\right] =0,
\end{array}
\label{D.3}
\end{equation}

\noindent where use was made of Eq.(\ref{2.2}).

The second term of Eq.(\ref{D.2}) is

\begin{equation}
-\int\tilde{dk} I(k) \left( k^i X^j - k^j X^i\right) = -\left( P^i X^j - P^j
X^i\right),  \label{D.4}
\end{equation}

\noindent where there is not a contribution from $X^0$ since $\omega (k)$ is
scalar.

Finally, the first term, using the definitions of $\hat I(k)$ and $\hat \phi
(k)$ can be written

\begin{equation}
\int\tilde{dk}\left( D {\cal H}(k)\right) \left( k^i{\frac{\partial}{%
\partial k^j}} - k^j{\frac{\partial}{\partial k^i}}\right) \hat{\phi}(k) =
\int\tilde{dk} {\cal H}(k) \left( k^i{\frac{\partial}{\partial k^j}} - k^j{%
\frac{\partial}{\partial k^i}}\right) {\cal K}(k),  \label{D.5}
\end{equation}

\noindent since $D$ is self-adjoint and commutes with $l_{ij}$ of Eq.(\ref
{3.15})

Collecting these terms we get the decomposition given in Eq.(\ref{6.1}).

An analogous calculation gives the decomposition (\ref{6.4}).

\vfill

\end{document}